\documentclass[preprint,aps,amsmath,nofootinbib]{revtex4-1}
\usepackage{graphicx,array,dcolumn}
\usepackage{calc,tabularx, epsfig,mathrsfs}
\usepackage{hyperref}

\usepackage{amsmath,verbatim,enumerate}
\usepackage{amssymb}
\usepackage{wasysym}
\usepackage{xcolor}
\usepackage{multirow}
\usepackage{xspace}
\usepackage{slashed}
\usepackage{mathtools}
\usepackage{cancel}
\allowdisplaybreaks[1]
\newlength{\figurewidth}
\newcommand{\beq}{\begin{equation}}
\newcommand{\eeq}{\end{equation}}
\newcommand{\bea}{\begin{eqnarray}}
\newcommand{\eea}{\end{eqnarray}}
\newcommand{\ba}{\begin{array}}
\newcommand{\ea}{\end{array}}

\newcommand{\mn}{{\mu\nu}}

\newcommand{\pt}{\partial}

\newcommand{\al}{\alpha}
\newcommand{\bt}{\beta}

\newcommand{\ep}{\epsilon}
\newcommand{\ta}{\theta}
\newcommand{\lam}{\lambda}
\newcommand{\Lam}{\Lambda}
\newcommand{\G}{\Gamma}

\newcommand{\de}{\delta}
\newcommand{\D}{\Delta}
\newcommand{\OM}{\Omega}

\newcommand{\sg}{\sigma}

\makeatother
\begin{document}
%
%
%
\title{
Lorentzian Robin Universe
}
\setlength{\figurewidth}{\columnwidth}
%
\author{Manishankar Ailiga}
\email{manishankara@iisc.ac.in}

\author{Shubhashis Mallik} 
\email{shubhashism@iisc.ac.in}

\author{Gaurav Narain}
\email{gnarain@iisc.ac.in}

\affiliation{
Center for High Energy Physics, Indian Institute of Science,
C V Raman Road, Bangalore 560012, India.
}

\vspace{5mm}

%
\begin{abstract}
In this paper, we delve into the gravitational path integral of 
Gauss-Bonnet gravity in four spacetime dimensions, in the 
mini-superspace approximation. Our primary focus lies in investigating 
the transition amplitude between distinct boundary configurations. 
Of particular interest is the case of Robin boundary conditions, 
known to lead to a stable Universe in Einstein-Hilbert gravity, alongside
Neumann boundary conditions. To ensure a consistent variational problem, 
we supplement the bulk action with suitable surface terms. This study leads 
us to compute the necessary surface terms required for Gauss-Bonnet gravity 
with the Robin boundary condition, which wasn't known earlier.
Thereafter, we perform an exact computation of the transition amplitude. 
Through $\hbar\to0$ analysis, we discover that the Gauss-Bonnet gravity 
inherently favors the initial configuration, aligning with the 
Hartle-Hawking no-boundary proposal. 
Remarkably, as the Universe expands, it undergoes a transition 
from the Euclidean (imaginary time) to the Lorentzian signature (real time). 
To further reinforce our findings, we employ a saddle point analysis 
utilizing the Picard-Lefschetz methods. The saddle point analysis 
allows us to find the initial configurations which lead to 
Hartle-Hawking no-boundary Universe that agrees with the 
exact computations. 
Our study concludes that for positive Gauss-Bonnet coupling, 
initial configurations corresponding to the Hartle-Hawking 
no-boundary Universe gives dominant contribution in the gravitational path-integral.
\end{abstract}


\maketitle

\tableofcontents


\section{Introduction}
\label{intro}

Transition amplitudes give us information about whether a particular 
transition is allowed. A change of endpoints (also known as boundary) will affect
the transition amplitude. Methods of quantum field theory (QFT) developed over the decades 
allow us to compute such transition amplitudes systematically, which can 
be matched with relevant experiments for verification. In fact, several experimental 
verifications of Standard Model predictions further strengthen 
our reliability on QFT and methods used to formulate it. Path integral
is one way to reliably study QFT and compute transition amplitudes
systematically.

However, to properly define path integral in QFT one needs to address
the issue of infinities that arise while computing transition amplitudes. 
These are taken care of by proper regularization and renormalization
\footnote{In gauge theories, one also has to 
do gauge-fixing to prevent over-counting of gauge degree of freedom 
and also suitably introduce relevant ghosts which ensures the gauge-fixing 
process is systematically done.}. Besides these, one still has to specify an 
integration contour as the standard Lorentzian path integrals 
are highly oscillatory and not absolutely convergent. In standard 
flat spacetime QFT (non-gravitational) this is achieved by
Wick rotation: a process of going from real-time to imaginary time. 
This transforms the original 
flat spacetime Lorentzian path integral into a convergent Euclidean 
path integral. 

These successes are hard to replicate when gravity is involved,
where besides dealing with issues of divergences, 
renormalizability and gauge-fixing, one also has to address the issue of 
the contour of integration. For the gravitational system, this need not be
standard Wick-rotation! 
Moreover, Euclidean gravitational path integral suffers from 
the conformal factor problem (the path integral over the conformal 
factor is unbounded from below \cite{Gibbons:1978ac}), 
motivating to study the gravitational path integral directly 
in the Lorentzian signature. A gravitational 
path integral with metric as the degree of freedom,
on a manifold with boundaries can be written as 
\beq
\label{eq:PI_LI_g}
G = \int_{{\cal M} + \pt {\cal M}} {\cal D} g_\mn e^{iS[g_\mn]/\hbar} \, ,
\eeq
where $g_\mn$ is the metric and $S[g_\mn]$ is the corresponding 
action. The path integral is defined on manifold ${\cal M}$ with boundary 
$\pt{\cal M}$. This abstract mathematical expression represents summation over 
all possible metrics with specified boundary conditions, 
where each geometry comes with a 
corresponding `weight-factor'.

The gravitational action that we will focus on in this paper
is given by following (see \cite{Narain:2021bff, Narain:2022msz}
for earlier works investigating the path integral of such 
gravitational theories)
\bea
\label{eq:act}
S = \frac{1}{16\pi G} \int {\rm d}^Dx \sqrt{-g}
\biggl[
-2\Lam + R + \al
\biggl( R_{\mu\nu\rho\sg} R^{\mu\nu\rho\sg} - 4 R_\mn R^\mn + R^2 \biggr)
\biggr] \, . 
\eea
This is the Gauss-Bonnet gravity action
where $G$ is the Newton's gravitational constant, 
$\Lam$ is the cosmological constant term,  
$\al$ is the Gauss-Bonnet (GB) coupling and $D$ is spacetime dimensionality. 
The mass dimensions of various couplings are: 
$[G] = M^{2-D}$, $[\Lam] = M^2$ and $[\al] = M^{-2}$. 

The action in eq. (\ref{eq:act}) falls in the class of lovelock gravity theories 
\cite{Lovelock:1971yv,Lovelock:1972vz,Lanczos:1938sf}, and
is a sub-class of higher-derivative gravity.
In this, the dynamical evolution equation of the metric 
field remains second order in time always. 
Interestingly, the GB term also arises in the
low-energy effective action of the heterotic string theory 
with $\al>0$ 
\cite{Zwiebach:1985uq,Gross:1986mw,Metsaev:1987zx}.
It should also be emphasized that for the first time, the GB-coupling $\al$ 
has received observational constraints
\cite{Chakravarti:2022zeq}. These constraints arise from the analysis 
of the gravitational wave (GW) data of the event 
GW150914 which also offered the first observational 
confirmation of the area theorem \cite{Isi:2020tac}.

Our interest in this paper is to study the path integral given in the 
eq. (\ref{eq:PI_LI_g}) for the gravitational action specified by
eq. (\ref{eq:act}), and to analyze the effects of the boundary conditions 
on the transition amplitude
(see 
\cite{York:1986lje,Brown:1992bq,Krishnan:2016mcj,
Witten:2018lgb,Krishnan:2017bte} for the role played by 
boundary terms).
We start by considering a 
spatially homogeneous and isotropic metric in $D$ spacetime 
dimensions. It is the FLRW metric in arbitrary spacetime 
dimension with dimensionality $D$. In polar
coordinates $\{t_p, r, \ta, \cdots \}$ it is given by
\beq
\label{eq:frwmet}
{\rm d}s^2 = - N_p^2(t_p) {\rm d} t_p^2 
+ a^2(t_p) \left[
\frac{{\rm d}r^2}{1-kr^2} + r^2 {\rm d} \OM_{D-2}^2
\right] \, .
\eeq
It has two unknown time-dependent functions: lapse $N_p(t_p)$
and scale-factor $a(t_p)$, $k=(0, \pm 1)$ is the curvature, 
and ${\rm d}\OM_{D-2}$ is the 
metric for the unit sphere in $D-2$ spatial dimensions. 
This is the \textit{mini-superspace} approximation of the metric.
In this approximation, though we don't have gravitational waves 
we still retain diffeomorphism invariance of the time 
co-ordinate $t_p$ and the dynamical scale-factor $a(t_p)$. 
This simple setting is enough to highlight the importance 
of boundary conditions and/or Gauss-Bonnet 
gravitational terms that might become relevant 
at the boundary. 

The Feynman path integral with reduced degree of freedom becomes
\beq
\label{eq:Gform_sch_fpt}
G[{\rm Bd}_f, {\rm Bd}_i]
= \int_{{\rm Bd}_i}^{{\rm Bd}_f} {\cal D} N_p {\cal D} \pi {\cal D} 
a(t_p) {\cal D} p {\cal D} {\cal C} {\cal D} \bar{P}
\exp \biggl[
\frac{i}{\hbar} \int_0^1 {\rm d} t_p \left(
N^\prime_p \pi + a^\prime p +{\cal C}^\prime \bar{P} - N_p H \right)
\biggr] \, ,
\eeq
where beside the integration over the scale-factor $a(t_p)$, lapse $N_p(t_p)$ 
and the Fermionic ghost ${\cal C}$, we also have an integration over 
their corresponding conjugate momenta given by
$p$, $\pi$, and $\bar{P}$ respectively.  
$({}^\prime)$ here denotes derivative 
with respect to $t_p$, while the time $t_p$ co-ordinate 
is chosen to range from $0\leq t_p \leq 1$. 
${\rm Bd}_i$ and ${\rm Bd}_f$ refers to the field configuration at 
initial ($t_p=0$) and final ($t_p=1$) boundaries respectively. 
The Hamiltonian constraint $H$ consists of two parts
\beq
\label{eq:Htwo}
H = H_{\rm grav}[a, p] + H_{\rm gh} [N, \pi, {\cal C}, \bar{P}] \, ,
\eeq
where $H_{\rm grav}$ refers to the Hamiltonian for the gravitational action 
and the Batalin-Fradkin-Vilkovisky (BFV) \cite{Batalin:1977pb}
ghost Hamiltonian is denoted by $H_{\rm gh}$
\footnote{ 
The BFV ghost is a generalization of the standard Fadeev-Popov ghost 
based on BRST symmetry. Usual gauge theories  
constraint algebra forms a Lie algebra. However, 
for gravitational systems respecting diffeomorphism invariance the 
constraint algebra doesn't close, requiring BFV quantization. 
In mini-superspace approximation, although the algebra 
trivially closes as there is only one constraint (Hamiltonian $H$)
but still, BFV quantization is preferable.}.
Although the degrees of freedom are quite reduced in mini-superspace 
approximation, the theory still retains time reparametrization invariance
which need to be gauge-fixed (we choose proper-time gauge $N^\prime_p=0$). 
For more elaborate discussion on BFV quantization process and ghost
see \cite{Teitelboim:1981ua,Teitelboim:1983fk,Halliwell:1988wc}.

Most of the path integral in eq. (\ref{eq:Gform_sch_fpt}) in the mini-superspace approximation
can be performed exactly (the path integral over ghosts (${\cal C}$ 
and $\bar{P}$) and the
conjugate momenta ($\pi$ and $p$)) 
leaving behind the following path integral 
\beq
\label{eq:Gform_sch}
G[{\rm Bd}_f, {\rm Bd}_i]
= \int_{0^+}^{\infty} {\rm d} N_p
\int_{{\rm Bd}_i}^{{\rm Bd}_f} {\cal D} a(t_p) \,\,
e^{i S[a, N_p]/\hbar} \, .
\eeq
The path integral $\int {\cal D} a(t_p) \,\, e^{i S[a, N_p]/\hbar}$
gives the transition amplitude for the
Universe to evolve from one boundary configuration to another
in the proper time $N_p$, while the lapse integration 
implies that one needs to consider paths of 
every proper duration $0<N_p<\infty$. 
This choice leads to causal evolution from 
one boundary configuration ${\rm Bd}_i$ to another ${\rm Bd}_f$ 
\cite{Teitelboim:1983fh}. 

The purpose of this paper is to study the path integral in 
eq. (\ref{eq:Gform_sch}) for the gravitational action given 
in eq. (\ref{eq:act}) focusing mainly on the case of Robin boundary condition (RBC)
imposed at the initial boundary. 
Imposing RBC at the initial time is like specifying a combination of 
field and its corresponding conjugate momenta. 
This is a broad class of RBC. In this paper, we focus on a subclass where RBC is imposed as a 
linear combination of field and its conjugate momenta. This translates into a one-parameter family of all allowed possible boundary conditions for a given 
specific value of the combination of field and its conjugate momenta.
Such a boundary condition interpolates between 
Dirichlet BC and Neumann BC.
Past studies dealing with Dirichlet BC showed that 
imposition of DBC at initial times 
leads to unsuppressed behavior of the gravitational fluctuations
in the no-boundary proposal of the Universe
\cite{Feldbrugge:2017kzv,DiTucci:2018fdg,Lehners:2018eeo,Feldbrugge:2017fcc,Feldbrugge:2017mbc},
which, with the DBC, corresponds to defining 
path-integral starting with zero size (see \cite{Lehners:2023yrj}
for review on the no-boundary proposal).
This study motivates one to investigate the situation in the 
case of Neumann and Robin BC at the initial boundary,
besides also indicating that Dirichlet BC is perhaps not the right boundary 
condition for gravity \cite{Krishnan:2016mcj,Krishnan:2017bte}.

Neumann boundary condition (NBC) has been studied 
in \cite{DiTucci:2019bui,Narain:2021bff, Lehners:2021jmv, DiTucci:2020weq, Narain:2022msz, Mondal:2023cxx}
using Picard-Lefschetz methods, where it was seen that 
gravitational fluctuations are well-behaved in the 
no-boundary proposal of the Universe,
which for Neumann BC is defined as the 
path-integral over regular geometries.
 Exact computations are done for the NBC case for the 
Gauss-Bonnet gravity further showed 
some surprises mentioned in \cite{Narain:2022msz}.
For Einstein-Hilbert gravity, Robin boundary condition
has been studied using the Picard-Lefschetz methods 
\cite{DiTucci:2019dji,DiTucci:2019bui,Mondal:2023cxx,Matsui:2023hei,Mondal:2022gyp}. These 
perturbative studies showed that 
fluctuations are well-behaved for the no-boundary 
proposal of the Universe.

In this paper, we take a fresh look at the implications 
of imposing RBC at the initial time and studying the 
path integral for the Gauss-Bonnet gravity. 
We study the Gauss-Bonnet gravity path integral using 
both perturbative and non-perturbative
methods to gain a proper understanding of the effects 
of RBCs on the behavior of the transition amplitude.
In the process, we also construct the suitable surface term 
for the Robin case needed for the Gauss-Bonnet gravity 
to have a consistent variational problem. We use time-slicing 
method of evaluating the gravitational path integral to compute the 
transition amplitude exactly. This is then compared with the
results obtained via Picard-Lefschetz (PL) methods to gain 
a better understanding of the role played by saddle 
point geometries which could be complex. 

PL methodology generalizes the notion of `Wick-rotation' by
adapting it to tackle 
highly oscillatory integral in a systematic manner 
\footnote{
Some studies involving Wick-rotation in curved spacetime 
have been initiated in \cite{Candelas:1977tt,Visser:2017atf,Baldazzi:2019kim,Baldazzi:2018mtl}.
However, more work needs to be done on this.
}. By providing a framework that takes into account contributions from all 
the possibly relevant complex saddle points of the path integral, 
this framework uniquely determines
contours of integration along which the oscillatory integrals 
(like the 
ones appearing in eq. (\ref{eq:Gform_sch})) becomes well-behaved.
Such contours termed \textit{Lefschetz thimbles}
constitute the generalized Wick-rotated contour.
In the context of Lorentzian quantum cosmology, these methods have been 
extensively used to analyze the nature of path integral 
\cite{Feldbrugge:2017kzv,Feldbrugge:2017fcc,Feldbrugge:2017mbc,
Vilenkin:2018dch, Vilenkin:2018oja, Rajeev:2021xit}
and the role played by boundaries 
\cite{DiTucci:2019dji,DiTucci:2019bui,
Narain:2021bff,Lehners:2021jmv,Narain:2022msz}
\footnote{
Earlier work using complex analysis methods to analyze 
Euclidean gravitational path integral was done in 
\cite{Hawking:1981gb,Hartle:1983ai}. Exploration of 
boundary effects in the Euclidean quantum cosmology
was done in 
\textit{tunnelling} proposal \cite{Vilenkin:1982de,Vilenkin:1983xq,Vilenkin:1984wp} 
and 
\textit{no-boundary} proposal \cite{Hawking:1981gb,Hartle:1983ai,Hawking:1983hj}.
Furthermore, the choice of a contour of integration via usage of 
complex analysis methods was also done in 
\cite{Halliwell:1988ik,Halliwell:1989dy,Halliwell:1990qr}.
}. 

PL-methods being inherently based on saddle-point approximation,
fall short of analyzing situations dealing with degenerate 
case {\it i.e.} when saddles become degenerate. In such 
cases, one has to go beyond saddle-point approximation,
as the approximation breaks down in such limiting cases. 
Non-perturbative and exact results whenever available 
become useful in investigating these situations.
This was particularly noticed in the study 
of gravitational path integral for the NBC case \cite{Narain:2022msz},
where it was seen that our Universe undergoes a transition 
from an Euclidean to a Lorentzian signature as the size of the Universe increases. 
However, saddle-point analysis broke down at the turning point,
a clear understanding of which came from the exact results. 
In the RBC case, it is expected that a similar situation 
could arise. It is therefore, best to approach the investigations 
of the gravitational path integral in the RBC case via both 
perturbative and non-perturbative manner.

The outline of the paper is as follows: section \ref{intro} gives 
introduction and motivation to the problem being studied. 
Section \ref{one-part} studies path integral
of a particle in potential in one dimension potential with various boundary conditions. 
This not only helps us understand path integrals with 
non-trivial boundary conditions but also give rise to relations 
between them, which becomes useful later in the paper.
Section \ref{minisup} discusses the mini-superspace 
approximation and writes the gravitational action in the 
mini-superspace approximation. 
Section \ref{bound_act} studies the variational problem 
and computes the surface terms that are needed to have a 
consistent variational problem. 
Section \ref{trans} studies the transition amplitude for the 
NBC and RBC case, and compute the exact expression for 
it by exploiting the results of section \ref{one-part}.
Section \ref{hbar0} studies the classical limit ($\hbar\to0$)
of the exact transition amplitude which shows the boundary 
configuration giving a dominant contribution in the path integral. 
Section \ref{sad_pot_approx} does the Picard-Lefschetz 
analysis of the contour integral. We conclude the paper 
in section \ref{conc} along with an outlook.

\section{One-dimensional Quantum Mechanics}
\label{one-part}

Before we proceed further with our study of gravitational path integral, we do a quick 
review of the path integral of one particle system to which our gravitational path integral 
reduces in the mini-superspace approximation. 

We look at the path integral of  one-particle system whose initial and final 
boundary conditions are specified. As the dynamical equation of motion 
involves two-derivatives of time, it needs only two boundary conditions. 
In the following, we will consider the cases where the final is always Dirichlet 
boundary condition and initial could be either Dirichlet or Neumann or Robin boundary 
condition. We will study a free-particle system and particle 
living in a linear potential which is of relevance to our studies in 
quantum cosmology.

Our purpose is to compute the following path integral
\beq
\label{1partPI}
\bar{G}[{\rm Bd_f}, {\rm Bd_i}] =
\int_{\rm Bd_i}^{\rm Bd_f} {\cal D} q(t) \,\, e^{i S_{\rm tot}[q]/\hbar} \, ,
\eeq
where $S_{\rm tot}[q]$ is the action for the one particle system 
\beq
\label{act1part}
S_{\rm tot}[q] = S[q] + S_{\rm bd} = 
\int_0^1 {\rm d}t \left[ \frac{m}{2} \dot{q}^2 - V(q) \right]
+ S_{\rm bd} \, ,
\eeq
$m$ is a $t$-independent parameter, 
`dot' denotes $t$-derivative, $V(q)$ is the potential,  
$S_{\rm bd}$ is the surface term added to have a 
consistent variational problem
\footnote{Variation of the action with respect to $q(t)$ to first 
order leads to terms proportional to $\de q(t)$ and its derivatives.
This will give rise to the equation 
of motion for $q(t)$ (terms proportional to $\de q(t)$), while any residual term will either 
vanish due to choice of boundary condition or has to be 
canceled by a suitable addition of a surface term}. 
The `bar' over $G$ in eq. (\ref{1partPI}) is added to 
prevent confusion later on with gravitational transition amplitude. 

The path integral in eq. (\ref{act1part}) can be 
computed by time-slicing method (first principles). The only subtlety arises 
at the end points around which the computation 
has to be done with care. To achieve this, we approach 
the problem in the {\it Hamiltonian} language as it is 
easier to incorporate the boundary conditions. 
The classical Hamiltonian for the above system is given by
\beq
\label{eq:Ham}
H(p,q) = p \dot{q} - L (q, \dot{q}) 
= \frac{p^2}{2m} + V(q) \, ,
\eeq
where $p = m \dot{q}$. The quantum transition amplitude 
from one state to another in the Hamiltonian picture 
(which is equivalent to computing the above path integral) 
is given by,
\beq
\label{eq:transH}
\bigl\langle {\rm Bd_f}, t=1 \big| {\rm Bd_i}, t=0 \bigr\rangle
= \bigl\langle {\rm Bd_f} \big| e^{-i H(\hat{p}, \hat{q})/\hbar} \big| {\rm Bd_i} \bigr\rangle \, .
\eeq
Here we have lifted $p$ and $q$ to operators $\hat{p}$ and 
$\hat{q}$ respectively, which then obey the commutation relation
\beq
\label{eq:commpq}
\big[\hat{q},\hat{p} \big] = i \hbar \, .
\eeq
It should be mentioned that the effect of surface terms $S_{\rm bd}$ gets taken 
care in the Hamiltonian style of doing the computation 
from one boundary state to another. The Hamiltonian follows 
from the Lagrangian by Legendre transform.

\subsection{Dirichlet boundary condition at $t=0$}
\label{ifDBC}

Let's first consider the case of a free particle where $V(q)=0$. The 
quantum Hamiltonian is given by
\beq
\label{eq:Hfree}
H_{\rm free} = \frac{\hat{p}^2}{2m} \, .
\eeq
One is interested in computing the transition amplitude from one state to another.
However, these states are given by specifying the initial and final position:
$q(t=1)=q_f$ and $q(t=0)=q_i$. The path integral in eq. (\ref{1partPI})
becomes
\beq
\label{eq:G_dbc_free}
\bar{G}^{\rm free}_{\rm DBC} (q_f, q_i)
= \bigl \langle q_f \big | e^{-i H_{\rm free}/\hbar} \big | q_i \bigr \rangle \, ,
\eeq
This can be evaluated from first principles by the time-slicing method, which 
discretizes the problem. Breaking the time-length into $M$ segments 
gives the time duration of each segment to be $\ep = 1/M$. This means that 
$q(t) \,\, :\Rightarrow \,\, q_0=q_i, q_1, q_2, \cdots, q_k, \cdot, q_{M-1}, q_M=q_f$, 
with the end points $q_0=q_i$ and $q_M=q_f$ are fixed.
This implies that we can write the transition amplitude 
given in eq. (\ref{eq:G_dbc_free}) in the following discretized form
\bea
\label{G_dbc_dis}
\bar{G}^{\rm free}_{\rm DBC} (q_f, q_i) 
&&
= \lim_{M\to\infty, \ep\to0} \biggl(\prod_{k=1}^{M-1} \int_{-\infty}^{\infty} {\rm d}q_k \biggr)
\langle q_M \big| U_{M,M-1} \big| q_{M-1} \rangle
\langle q_{M-1} \big| U_{M-1,M-2} \big| q_{M-2} \rangle \cdots
\notag \\
&&
\times \langle q_k \big| U_{k,k-1} \big| q_{k-1} \rangle \cdots 
\langle q_2 \big| U_{2,1} \big| q_1 \rangle
\langle q_1 \big| U_{1,0} \big| q_0 \rangle \, ,
\eea 
where 
\beq
\label{eq:Uk}
U_{k,k-1} = e^{-i H_{\rm free} \ep /\hbar} .
\eeq
Consider the following matrix element 
$\langle q_k \big| e^{-i H_{\rm free} \ep/\hbar} \big|q_{k-1}\rangle$. In this, one can 
plug completeness-relation in momentum space and perform the momentum integral
\bea
\langle q_k \big| e^{-i H_{\rm free} \ep/\hbar} \big|q_{k-1}\rangle 
&&
= \int_{-\infty}^{\infty} \frac{{\rm d} p_k}{2\pi \hbar}  
\langle q_k \big| e^{-i H_{\rm free} \ep/\hbar} \big| p_k \rangle \langle p_k \big| q_{k-1}\rangle \, ,
\notag \\
&&
= \int_{-\infty}^{\infty} \frac{{\rm d} p_k}{2\pi \hbar} 
e^{
-i \ep p_k^2/(2m\hbar) +i p_k(q_k - q_{k-1})/\hbar
}
= \sqrt{\frac{m}{2\pi i \ep \hbar}} e^{i m (q_k - q_{k-1})^2/(2\ep\hbar)} ,
\eea
where we performed the Gaussian integral in $p_k$ 
to obtain the last expression. Eventually, we are left with
the following expression for $\bar{G}^{\rm free}_{\rm DBC}$
\beq
\label{eq:Gdbc_am}
\bar{G}^{\rm free}_{\rm DBC}(q_f, q_i) 
= \lim_{M\to\infty, \ep\to0} 
\biggl(\prod_{k=1}^{M-1} \int_{-\infty}^{\infty} {\rm d}q_k \biggr)
\left(\frac{m}{2\pi i \ep \hbar}\right)^{M/2}
\exp\left[\frac{im}{2\ep\hbar} \sum_{k=1}^M (q_k - q_{k-1})^2 \right] \, .
\eeq
The series of Gaussian $q$-integrals can be carried out one-by-one starting from $q_1$
till we reach integration over $q_{M-1}$
which is the final integration that one has to perform. This will lead 
to the following expression for $\bar{G}^{\rm free}_{\rm DBC}(q_f, t=1; q_i, t=0)$
\beq
\label{Gdbcfree}
\bar{G}^{\rm free}_{\rm DBC} (q_f, t=1; q_i, t=0) = 
\sqrt{\frac{m}{2\pi i \hbar}} e^{i m (q_f - q_i)^2/2\hbar} \, ,
\eeq 
where we have used $M \ep =1$, when taking the limit 
$M\to\infty$ and $\ep\to0$. In the case interaction term $V(q)$ is present,
then eq. (\ref{eq:Gdbc_am}) gets modified into the following
\bea
\label{eq:Gdbc_v}
\bar{G}_{\rm DBC}(q_f, q_i) = 
&&
\lim_{M\to\infty, \ep\to0} 
\biggl(\prod_{k=1}^{M-1} \int_{-\infty}^{\infty} {\rm d}q_k \biggr)
\left(\frac{m}{2\pi i \ep \hbar}\right)^{M/2}
\notag \\
&& \times
\exp\left[\frac{im}{2\ep \hbar} \sum_{k=1}^M (q_k - q_{k-1})^2 
- \frac{i \ep}{\hbar} V(q_{k-1}) \right] \, .
\eea
This is nothing but a discretized version of the path integral given in eq. (\ref{1partPI})
where $\dot{q}_{k} = (q_k - q_{k-1})/\ep$. However, for a generic potential 
$V(q)$ this can't be computed exactly. In the case for linear potential $V(q) = \lam q$ 
(which will also appear in quantum cosmology, as we will later see),
one can easily compute the above-discretized version of the path integral exactly.
The end result after the $q$-integrations is the following,
\bea
\label{eq:Gdbc_v_end}
\bar{G}_{\rm DBC}(q_f, q_i) &&= \sqrt{\frac{m}{2\pi i \hbar}} 
\exp\biggl[
\frac{i}{\hbar}\left\{\frac{m(q_f-q_i)^2}{2}- \frac{\lam(q_{f}+q_{i})}{2} - \frac{\lam^2}{24 m}
\right\}
\biggr]\, ,
\notag \\
&&= \bar{G}^{\rm free}_{\rm DBC}(q_f, q_i) \,\,
e^{-i\lam \left\{
\lam + 12 m (q_f + q_i)
\right\}/(24m\hbar)}\, .
\eea
Notice the quadratic structure of the $\lam$-dependent terms.
The part coming from interaction vanishes for two values of 
$\lam$: $\lam=0$ (free theory) and $\lam= - 12 m(q_f+q_i)$ (interacting).

\subsection{Neumann boundary condition at $t=0$}
\label{NBCt0_free}

The path integral for the Neumann boundary condition is easier to deal with 
when handled in the Hamiltonian framework. The quantity we are interested in 
computing is
\beq
\label{eq:Gnbc}
\bar{G}_{\rm NBC} (q_f, t=1; p_i, t=0) = \langle q_f \big | e^{-i H/\hbar} \big| p_i \rangle \, ,
\eeq
where at the initial time, we fix the momentum.
We note that the initial momentum state can be written as the following
\beq
\label{PinQ}
\big| p_i \rangle = \int_{-\infty}^{\infty} {\rm d}q_0 
\,\, e^{i p_i q_0/\hbar} \,\, \big| q_0 \rangle \, .
\eeq
This allows one to express $\bar{G}_{\rm NBC}$ as a Fourier transform of 
another transition amplitude 
\beq
\label{eq:Gnbc_f}
\bar{G}_{\rm NBC}(q_f, t=1; p_i, t=0) = \int_{-\infty}^{\infty} {\rm d}q_0 
\,\, e^{i p_i q_0/\hbar} \,\, 
\underbracket{\langle q_f \big | e^{-i H/\hbar} \big| q_0 \rangle}_{\bar{G}_{\rm DBC}(q_f, t=1; q_0, t=0)} \, .
\eeq
This shows that $\bar{G}_{\rm NBC}(q_f, t=1; p_i, t=0)$ is related 
to $\bar{G}_{\rm DBC}(q_f, t=1; q_0, t=0)$ by a Fourier transform
thereby implying that one can be obtained from another 
by transformation (if $\bar{G}_{\rm NBC}(q_f, t=1; p_i, t=0)$
is known then $\bar{G}_{\rm DBC}(q_f, t=1; q_0, t=0)$ can be obtained from 
it by an inverse Fourier transform). In the present 
situation as we have already worked out the transition 
amplitude $\bar{G}_{\rm DBC}(q_f, t=1; q_0, t=0)$, the above Fourier 
transform can be used to compute the expression for the 
Neumann transition amplitude. This is given by
\beq
\label{eq:Gnbc_free}
\bar{G}^{\rm free}_{\rm NBC}(q_f, t=1; p_i, t=0)
=  e^{i(p_{i}q_{f}-p_{i}^2/2m)/\hbar} \, .
\eeq 
Similarly, one can exploit the above 
technique to work out the path integral for a particle in the linear-potential 
with Neumann boundary condition at the initial time. 
Here if we plug eq. (\ref{eq:Gdbc_v})
on the RHS of eq. (\ref{eq:Gnbc_f}) and writing $q_i \to q_0$, then 
on integration with respect to $q_0$ we get the $\bar{G}_{\rm NBC}(q_f, p_i)$. 
\bea
\label{Gnbc_v}
\bar{G}_{\rm NBC}(q_{f}, p_{i})  &=& 
\exp \biggl[
\frac{i}{\hbar} \left\{
p_{i}q_{f}- \frac{p_{i}^2}{2m} 
- \frac{\lam \left(\lam - 3 p_{i} + 6m q_{f}\right)}{6m}
\right\}
\biggr] \, ,
\notag \\
&=& \bar{G}^{\rm free}_{\rm NBC}(q_{f}, p_{i}) \times
e^{-i \lam \left(\lam - 3 p_{i} + 6m q_{f}\right)/(6m\hbar)} \, .
\eea
The interaction dependent term is quadratic in $\lam$ and vanishes
for two values of $\lam$: $\lam=0$ and $\lam = 3(p_{i}- 2 m q_{f})$.

\subsection{Robin boundary condition at $t=0$}
\label{rbc_free}

Now we are interested in computing the path integral for the case 
when Robin boundary condition (RBC) is specified at the initial time $t=0$. 
An RBC at $t=0$ means that a given linear combination of 
position and momentum is fixed at that time. 

The boundary value problem we are interested in is the following
\beq
\label{eq:rbc}
{\rm Bd_f}: q ({t=1})= q_{f}  \quad \text{and} \quad {\rm Bd_i}: p_{i}+\bt \,q_{i} = P_i \, ,
\eeq
where $q_{i}\equiv q(t=0)$ and $p_{i}\equiv p(t=0)$. The path integral we are 
interested in computing is
\beq
\label{eq:Grbc-def}
\bar{G}_{\rm RBC} (q_f, t=1; p_{i}+\bt \,q_{i}, t=0)
= \langle q_f \big| e^{-i H/\hbar} \big| p_{i}+\bt \,q_{i} \rangle \, .
\eeq
An interesting way to evaluate this is to do a canonical transformation and 
define a new momentum variable $P$ and position variable $Q$ as 
\beq
\label{eq:newP}
P(t) = p(t) + \bt q(t) 
\quad {\rm and} \quad
Q(t) = q(t) \, .
\eeq
Under this transformation, the commutation relation is preserved, i.e.,
\beq
\label{C}
[\hat{Q}, \hat{P}]=[\hat{q}, \hat{P}]=[\hat{q}, \hat{p}+\bt \hat{q}]=i \hbar \, .
\eeq
This also means that the Jacobian of transformation is unity.
However, in terms of new variables, the Hamiltonian acquires a new form
$H(\hat{p},\hat{q})=H(\hat{P} - \bt \hat{Q}, \hat{Q}) \equiv H_\bt(\hat{P}, \hat{Q})$ 
(where subscript `$\bt$' symbolizes $\bt$-dependence of the Hamiltonian).
In terms of new variables, our transition amplitude acquires the 
following form
\beq
\label{Gnrbc}
\bar{G}_{\rm RBC} (q_f, t=1; p_{i}+\bt \,q_{i}, t=0)
= \bar{G}_{\rm NBC}  (Q_f, t=1; P_i, t=0) 
= \langle Q_f \big| e^{-i H_\bt /\hbar} \big| P_i \rangle \, .
\eeq
In terms of new variables the Robin boundary condition 
problem transforms into an NBC problem.
One can write the initial state as follows  
\beq
\label{PiF}
\big| P_i \rangle =
\int_{-\infty}^{\infty} {\rm d} Q_0 \, \, e^{i P_i Q_0/\hbar} \,\, \big| Q_0 \rangle \, .
\eeq
This means that for the Robin boundary condition the path integral 
becomes the following
\beq
\label{Grbc_four}
\bar{G}_{\rm RBC} (q_f, t=1; p_{i}+\bt \,q_{i}, t=0)
= \int_{-\infty}^{\infty} {\rm d} Q_0 \, \, e^{i P_i Q_0/\hbar}
\underbracket{\langle Q_f \big| e^{-i H_\bt(\hat{P},\hat{Q})/\hbar}\big| Q_0 \rangle}
_{\bar{G}^\bt_{\rm DBC}(Q_f, t=1; Q_0, t=0)} \, .
\eeq
The object of interest here is $\langle Q_f \big| e^{-i H_\bt(\hat{P},\hat{Q})/\hbar}\big| Q_0 \rangle$
which is the DBC path integral for the transformed Hamiltonian $H_\bt(P,Q)$.
It should be pointed out that this process of evaluating 
RBC path integrals remain valid even for interacting theories, 
but to gain clarity on the methods, we first consider free theory.
The free Hamiltonian $H_\bt^{\rm free}$ is given by
\beq
\label{freeHPQ}
H^{\rm free}_\bt (\hat{P},\hat{Q}) = \frac{(\hat{P}-\bt \hat{Q})^2}{2m}
= \frac{\hat{P}^2}{2m}+\frac{\bt^2 \hat{Q}^2}{2m}
-\frac{\bt}{2m}(\hat{Q} \hat{P} + \hat{P} \hat{Q}) \, ,
\eeq
and in transformed variables the boundary conditions become
\beq
\label{BC_rbc_tr}
Q(t=1) = Q_{f} = q_f \quad \text{and} \quad P(t=0) = P_i \, .
\eeq
Therefore, our original interests in computing path integral with RBC 
shifts to computing path integral in new variables but with NBC. However, the 
Hamiltonian in new variables is different which means that one needs to be careful 
while computing path integral via time-slicing method.
We are interested in computing something like eq. (\ref{G_dbc_dis})
with the Hamiltonian given in eq. (\ref{freeHPQ}). Let us consider the 
following matrix 
\bea
\label{eq:matUn}
&&
\langle Q_{k+1} \big| U_{k+1, k} \big| Q_k \rangle
= \int_{-\infty}^{\infty} \frac{{\rm d} P_k}{2\pi \hbar} 
\langle Q_{k+1} \big|
e^{-i \ep H^{\rm free}_\bt(\hat{P}, \hat{Q})/\hbar} \big| P_k \rangle\langle P_k \big| Q_k \rangle \, ,
\notag \\
&&
= \left(\frac{m}{2\pi i \ep \hbar} \right)^{1/2} e^{\bt \ep/2m} \,\,\,\,
\exp\biggl[\frac{i}{\hbar}
\left\{\frac{m (Q_{k+1} - Q_{k})^2}{2\ep} + \bt Q_{k+1} (Q_{k+1} - Q_k)
\right\}
\biggr] \, .
\eea
This means that 
\bea
\label{eq:Grbc_ev}
\bar{G}^{\rm free}_{\rm RBC} (Q_f, t=1; && P_i, t=0)
= \lim_{M\to\infty, \ep\to0}
\left(\prod_{k=0}^{M-1} \int_{-\infty}^{\infty} {\rm d} Q_k
\left(\frac{m}{2\pi i \ep \hbar} \right)^{1/2} e^{\bt \ep/2m}
\right)
\notag \\
&&
\times
e^{i P_i Q_0/\hbar} 
e^{i\bt (Q_M^2 - Q_0^2)/(2\hbar)}
e^{im \sum_{k=0}^{M-1} (Q_{k+1} - Q_k)^2/(2\ep\hbar)} \, .
\eea
The term in the exponent $\bt(Q_M^2 - Q_0^2)/2$ is obtained by noticing that 
$\bt \sum_{k} Q_{k+1} (Q_{k+1} - Q_k)$ is actually a discretized 
version of $ \bt \int_0^1 {\rm d} t \, Q \dot{Q} $, which can
be written as a total derivative 
$ (\bt/2) \int_0^1 {\rm d} t \, {\rm d}Q^2/{\rm d}t $.
This eventually gives a simplified expression for the 
Robin-boundary condition transition amplitude.
\beq
\label{eq:Grbc_exp}
\bar{G}^{\rm free}_{\rm RBC} (Q_f, t=1; P_i, t=0)
=  \int_{-\infty}^{\infty} {\rm d} Q_0 \, \, e^{i P_i Q_0/\hbar} 
e^{i\bt (Q_M^2 - Q_0^2)/(2\hbar)} \,\,\,
\bar{G}_{\rm DBC}^{\rm free}[Q_M, t=1; Q_0, t=0] \, .
\eeq
This can be easily computed after plugging the value 
of $\bar{G}_{\rm DBC}^{\rm free}$ from eq. (\ref{Gdbcfree}) which gives
\beq
\label{Grbc_free_exp}
\bar{G}^{\rm free}_{\rm RBC} (Q_f, t=1; P_i, t=0)
= \sqrt{\frac{m}{m-\bt}}
\exp\biggl[
\frac{i}{2\hbar}
\left\{
\bt Q_f^2
- \frac{\bt m Q_f^2 + P_i^2 - 2 m P_i Q_f}{m-\bt} 
\right\}
\biggr]
\, ,
\eeq
which in limit $\bt \to0$ reproduces the expression for the 
Neumann boundary condition result given. in eq. (\ref{eq:Gnbc_free}). 

In the case when interactions are present, the modified 
Hamiltonian gets amended by potential term $V(Q)$ 
\beq
\label{eq:Hn_mod_V}
H_\bt (\hat{P},\hat{Q}) = \frac{(\hat{P}-\bt \hat{Q})^2}{2m} + V(\hat{Q}) \, .
\eeq
Then following the same steps as above one arrives at an analogous 
expression for $\langle Q_{k+1} \big| U_{k+1, k} \big| Q_k \rangle$
which is given by
\bea
\label{eq:matUn_V}
\langle Q_{k+1} \big| U_{k+1, k} \big| Q_k \rangle
=&& \left(\frac{m}{2\pi i \ep \hbar} \right)^{1/2} e^{\frac{\bt \ep}{2m}} 
\exp
\biggl[\frac{i}{\hbar}
\biggl\{
\frac{m (Q_{k+1} - Q_{k})^2}{2\ep} 
\notag \\
&&
+ \bt Q_{k+1} (Q_{k+1} - Q_k)
- \ep V(Q_k)
\biggr\}
\biggr] \, .
\eea
Once again the term $\bt \sum_k Q_{k+1} (Q_{k+1} - Q_k)$ is discretized 
version of $\bt \int_0^1 {\rm d} t \, Q \dot{Q} $ and can
be written as a total derivative 
$(\bt/2) \int_0^1 {\rm d} t \, {\rm d}Q^2/{\rm d}t $. Then the 
expression for the $\bar{G}_{\rm RBC}$ with 
the potential is given by
\bea
\label{eq:Grbc_ev_V}
&& \bar{G}_{\rm RBC} (Q_f, t=1; P_i, t=0)
= \lim_{M\to\infty, \ep\to0}
\left(\prod_{k=0}^{M-1} \int_{-\infty}^{\infty} {\rm d} Q_k
\left(\frac{m}{2\pi i \ep \hbar} \right)^{1/2} e^{\bt \ep/2m}
\right) 
\notag \\
&& 
\times
\exp\biggl[
\frac{i}{\hbar}
\left\{
P_i Q_0
+ \frac{\bt (Q_M^2 - Q_0^2)}{2}
+ \frac{m}{2\ep} \sum_{k=0}^{M-1} (Q_{k+1} - Q_k)^2 - \ep \sum_{k=0}^{M-1} V(Q_k)
\right\}
\biggr] \, ,
\notag \\
&&
=  \int_{-\infty}^{\infty} {\rm d} Q_0 \, \, e^{i P_i Q_0/\hbar} 
e^{i\bt (Q_M^2 - Q_0^2)/(2\hbar)} \,\,\,
\bar{G}_{\rm DBC}[Q_M, t=1; Q_0, t=0] \, .
\eea
It should be highlighted that this last line is valid for any arbitrary potential $V(Q)$.
For the case of linear potential one can plug the expression 
for $\bar{G}_{\rm DBC}[Q_M, t=1; Q_0, t=0]$ from the eq. (\ref{eq:Gdbc_v}) 
in the above to obtain
$\bar{G}_{\rm RBC} (q_f, t=1; p_{i}+\bt \,q_{i}, t=0)$
\bea
\label{Grbc_V_exp}
\bar{G}_{\rm RBC} (Q_f, t=1; P_i, t=0)
=&& \bar{G}^{\rm free}_{\rm RBC} (Q_f, t=1; P_i, t=0)
\notag \\
&& \times
\exp\left[
\frac{i\lam \left\{
\bt q_f +(P_i - 2 m q_f)
\right\}}{2(m-\bt)\hbar} 
- \frac{i\lam^2 (4m-\bt)}{24m(m-\bt)\hbar}
\right] \, .
\eea
Notice that in the limit $\bt\to0$ this RBC expression goes to the 
NBC expression given in eq. (\ref{Gnbc_v}). 
The full Robin path integral for interacting theory is a product of 
Robin path integral for free theory and a coupling dependent piece. 

The expression given in eq. (\ref{eq:Grbc_ev_V}) can be converted 
into a relation between $\bar{G}_{\rm RBC} (Q_f, t=1; P_i, t=0)$
and $\bar{G}_{\rm NBC}(q_f, t=1; p_{i}, t=0)$ by exploiting the inverse Fourier transform. 
After a bit of algebra and computation of $Q_0$-Gaussian integral, it is seen 
that 
\bea
\label{eq:Grbc_Gnbc}
\bar{G}_{\rm RBC} (Q_f, t=1; P_i, t=0)
= \sqrt{\frac{e^{i\bt Q_f^2}}{i \bt} }
\int_{-\infty}^{\infty} \frac{{\rm d} \tilde{p}}{\sqrt{2 \pi \hbar}}
e^{i\frac{(P_i - \tilde{p})^2}{2\bt\hbar}} \bar{G}_{\rm NBC}(q_f, t=1; \tilde{p}, t=0) \, ,
\eea
where $P_i = p_{i}+\bt \,q_{i}$. This is an exact relation connecting the 
two quantities. In the following, we will use this relation 
to compute the exact expression of gravitational path integral in mini-superspace
approximation for Robin BC and investigate it in detail to study the 
implications of Robin BC on the evolution of the Universe.

\section{Mini-superspace action}
\label{minisup}

We start by considering the FLRW metric given 
in eq. (\ref{eq:frwmet}) which is conformally-flat and hence
has a vanishing Weyl-tensor ($C_{\mu\nu\rho\sg} =0$). 
For a generic $d$-dimensional FLRW metric the non-zero 
entries of the Riemann tensor are 
\cite{Deruelle:1989fj,Tangherlini:1963bw,Tangherlini:1986bw} 
\bea
\label{eq:riemann}
R_{0i0j} &=& - \left(\frac{a^{\prime\prime}}{a} - \frac{a^\prime N_p^\prime}{a N_p} \right) g_{ij} \, , 
\notag \\
R_{ijkl} &=& \left(\frac{k}{a^2} + \frac{a^{\prime2}}{N_p^2 a^2} \right)
\left(g_{ik} g_{jl} - g_{il} g_{jk} \right) \, .
\eea
Here $g_{ij}$ is the spatial part of the FLRW metric
while $({}^\prime)$ denotes the derivative with respect to $t_p$.
The non-zero components of Ricci-tensor are
\bea
\label{eq:Ricci-ten}
R_{00} &=& - (D-1) \left(\frac{a^{\prime\prime}}{a} - \frac{a^\prime N_p^\prime}{a N_p} \right)
\, , 
\notag \\
R_{ij} &=& \left[
\frac{(D-2) (k N_p^2 + a^{\prime2})}{N_p^2 a^2}
+ \frac{a^{\prime\prime} N_p - a^\prime N_p^\prime}{a N_p^3} 
\right] g_{ij} \, ,
\eea
while the Ricci-scalar for FLRW is given by
\beq
\label{eq:Ricci0}
R = 2(D-1) \left[\frac{a^{\prime\prime} N_p - a^\prime N_p^\prime}{a N_p^3} 
+ \frac{(D-2)(k N_p^2 + a^{\prime2})}{2N_p^2 a^2} \right]
\, .
\eeq
Weyl-flat metrics also have a wonderful property that allows 
one to express the Riemann tensor in terms of Ricci-tensor and Ricci scalar as follows
\bea
\label{eq:Riem_exp}
R_{\mu\nu\rho\sg} = \frac{R_{\mu\rho} g_{\nu\sg} - R_{\mu\sg}g_{\nu\rho}
+ R_{\nu\sg} g_{\mu\rho} - R_{\nu\rho} g_{\mu\sg}}{D-2}
- \frac{R (g_{\mu\rho} g_{\nu\sg} - g_{\mu\sg} g_{\nu\rho})}{(D-1)(D-2)} \,.
\eea
This furthermore helps one to express the square of Riemann-tensor 
as a linear combination of Ricci-tensor-square and Ricci-scalar-square. 
This is given by
\beq
\label{eq:Reim2_exp}
R_{\mu\nu\rho\sg} R^{\mu\nu\rho\sg}
= \frac{4}{D-2} R_\mn R^\mn - \frac{2 R^2}{(D-1)(D-2)} \, .
\eeq
This identity is often used to simplify the expression for 
the Gauss-Bonnet gravity action for the case of 
Weyl-flat metrics.
\bea
\label{eq:actGB}
\int {\rm d}^Dx \sqrt{-g} && \left(
R_{\mu\nu\rho\sg} R^{\mu\nu\rho\sg}  - 4 R_\mn R^\mn + R^2
\right)
\notag \\
&&
= \frac{D-3}{D-2} \int {\rm d}^Dx \sqrt{-g} \left(
- 4 R_\mn R^\mn + \frac{D R^2}{D-1}
\right) \, .
\eea
On plugging the FLRW metric of eq. (\ref{eq:frwmet}) in the gravitational action 
stated in eq. (\ref{eq:act}), we get an action for scale-factor $a(t_p)$ and lapse
$N_p(t_p)$ in $D$-dimensions
\bea
\label{eq:midSact}
&&
S[a,N_{p}] = \frac{V_{D-1}}{16 \pi G} \int {\rm d}t_p
\biggl[
\frac{a^{D-3}}{N_p^2} \biggl\{
(D-1)(D-2) k N_p^3 - 2 \Lam a^2 N_p^3 - 2 (D-1) a a^\prime N_p^\prime
\notag \\
&&
+ (D-1)(D-2) a^{\prime2} N_p + 2 (D-1) N_p a a^{\prime\prime}
\biggr\}
+ (D-1)(D-2)(D-3) \al\biggl\{
\frac{a^{D-5}(D-4)}{N_p^3} 
\notag\\
&&
\times (kN_p^2 + a^{\prime2})^2 
+ 4 a^{D-4}\frac{{\rm d}}{{\rm d}t_p} 
\left(
\frac{k a^\prime}{N_p} + \frac{a^{\prime 3}}{3N_p^3}
\right)
\biggr\}
\biggr] \, .
\eea
$V_{D-1}$ here refers to the volume of the $D-1$ dimensional spatial hyper-surface
which is given by,
\beq
\label{eq:volDm1}
V_{D-1} = \frac{\G(1/2)}{\G(D/2)} \left(\frac{\pi}{k}\right)^{(D-1)/2} \, .
\eeq
In four spacetime dimensions ($D=4$) the terms proportional to $\al$
(Gauss-Bonnet coupling parameter)  
becomes a total time derivative. In $D=4$ the mini-superspace 
gravitational action then becomes the following
\beq
\label{eq:mini_sup_d4}
S[a,N_{p}] = \frac{V_{3}}{16 \pi G} \int {\rm d}t_p
\biggl[
6k a N_p - 2 \Lam a^3 N_p - \frac{6 a^2 a^\prime N_p^\prime}{N_p}
+ \frac{6 a a^{\prime 2}}{N_p} + \frac{6 a^{\prime\prime} a^2}{N_p}
+ 24 \al \frac{{\rm d}}{{\rm d}t_p} 
\left(
\frac{k a^\prime}{N_p} + \frac{a^{\prime 3}}{3N_p^3}
\right)
\biggr] \, .
\eeq
At this point, one can introduce rescaling of lapse $N_p$ and
scale factor to bring out a more appealing form of the 
mini-superspace action by doing the following transformation 
\beq
\label{eq:rescale}
N_p(t_p) {\rm d} t_p = \frac{N(t)}{a(t)} {\rm d} t \, ,
\hspace{5mm}
q(t) = a^2(t) \, .
\eeq
This co-ordinate transformation re-expresses our original 
FLRW metric in eq. (\ref{eq:frwmet}) in a different form.
\beq
\label{eq:frwmet_changed}
{\rm d}s^2 = - \frac{N^2}{q(t)} {\rm d} t^2 
+ q(t) \left[
\frac{{\rm d}r^2}{1-kr^2} + r^2 {\rm d} \OM_{D-2}^2
\right] \, .
\eeq
Notice also that in the new coordinate frame, the new `time'
is denoted by $t$.
Under this transformation, the original gravitational action in $D=4$ 
given in eq. (\ref{eq:mini_sup_d4}) 
acquires the following simple form
\bea
\label{eq:Sact_frw_simp}
S[q,N] = \frac{V_3}{16 \pi G} \int_0^1 {\rm d}t \biggl[
(6 k - 2\Lam q) N + \frac{3 \dot{q}^2}{2N}
+ 3q \frac{{\rm d}}{{\rm d} t} \left(\frac{\dot{q}}{N} \right)
+ 24 \al \frac{{\rm d}}{{\rm d} t} \left(
\frac{k\dot{q}}{2N} + \frac{\dot{q}^3}{24 N^3} 
\right)
\biggr] \, .
\eea
Here $(\dot{})$ represent time $t$ derivative. 
Note that the GB-term is a total time derivative.
Integration by parts of the terms in the action
re-expresses the action in a more recognizable form along with
some surface terms. 
\bea
\label{eq:Sact_frw_simp_inp}
S[q,N] &&
=\frac{V_3}{16 \pi G} \int_0^1 {\rm d}t \biggl[
(6 k - 2\Lam q) N - \frac{3 \dot{q}^2}{2N} \biggr] 
+ \frac{V_3}{16 \pi G} \biggl[
\frac{3q_f \dot{q}_f}{N} - \frac{3q_i \dot{q}_i}{N}
\notag \\
&&
+ 24 \al \left(
\frac{k\dot{q_f}}{2N} + \frac{\dot{q_f}^3}{24 N^3} 
- 
\frac{k\dot{q_i}}{2N} - \frac{\dot{q_i}^3}{24 N^3} 
\right)
\biggr]
\, .
\eea
The surface terms consist of two parts: one coming 
from EH-part of gravitational action while the other is from 
GB sector. The bulk term is like an action of a one-particle 
system in a linear potential.
From now onwards we will work with the convention 
that $V_3 = 8\pi G$, and in the rest of the paper we will study the 
path integral of this action.

\section{Action variation and Boundary terms}
\label{bound_act}

We start by constructing the variational problem by 
considering the variation of the action in eq. (\ref{eq:Sact_frw_simp}) 
with respect to $q(t)$. This will not only yield terms that will 
eventually lead to the equation of motion but also generate a 
collection of boundary terms. 

In the rest of the paper, we will work with the ADM gauge 
$\dot{N}=0$, which implies 
setting $N(t) = N_c$ (constant). To set up the variational 
problem consistently and avoid missing of any terms we write 
\beq
\label{eq:qfluc}
q(t) = \bar{q}(t) + \ep \de q(t) \, ,
\eeq
where $\bar{q}(t)$ is some background $q(t)$ and $\de q(t)$ is the fluctuation 
around this. The parameter $\ep$ is used to keep track of the order of fluctuation terms. 
On plugging this in the action given in eq. (\ref{eq:Sact_frw_simp}) and on 
expanding it to first order in $\ep$, we notice that the terms
proportional to $\ep$ are given by following
\beq
\label{eq:Sexp_qvar}
\de S = \frac{\ep}{2} \int_{0}^{1} {\rm d}t \biggl[
\left(-2 \Lam N_c + \frac{3 \ddot{\bar{q}}}{N_c} \right) \de q
+ \frac{3}{N_c} \frac{{\rm d}}{{\rm d} t} \left(\bar{q} \de \dot{q} \right)
+ 24 \al \frac{{\rm d}}{{\rm d} t} \left\{ 
\left(\frac{k}{2N_c} + \frac{\dot{\bar{q}}^2}{8N_c^3} \right) \de \dot{q} \right\}
\biggr] \, .
\eeq
We also notice that in the above expression, there are two total time-derivative pieces 
which contribute at the boundaries. These boundary pieces need to 
be canceled appropriately by supplementing the action with 
suitable surface terms in order to have a consistent 
boundary value problem for a particular choice of boundary condition.
The terms proportional to $\de q$ on the other hand 
will lead to the equation of motion if one demands that the full
expression multiplying it vanishes. 
This gives the dynamical equation obeyed by $\bar{q}(t)$
\beq
\label{eq:dyn_q_eq}
\ddot{\bar{q}} = \frac{2}{3} \Lam N_c^2 \, .
\eeq
This process of construing surface action is based 
on analyzing the behavior of fluctuations around the 
classical trajectory (on-shell geometries), and may sometime miss 
to properly capture the features of topological effects
as by definition topological terms don't play a role 
in the equation of motion. This will become more clear 
as we proceed further in the paper.

The second-order ODE given in eq(\ref{eq:dyn_q_eq}) is easy to solve 
and its general solution is given by
\beq
\label{eq:qsol_gen}
\bar{q}(t) = \frac{\Lam N_c^2}{3} t^2 + c_1 t + c_2 \, .
\eeq
The constants $c_{1,2}$ are determined by requiring the solution to 
satisfy the boundary conditions. 
The total-derivative terms in eq. (\ref{eq:Sexp_qvar}) 
lead to a collection of boundary terms
\beq
\label{eq:Sbd}
S_{\rm bdy} = \frac{\ep}{2} \biggl[
\frac{3}{N_c} \left(\bar{q}_f \de \dot{q}_f - \bar{q}_i \de \dot{q}_i \right)
+ 24 \al \left\{ 
\left(\frac{k\de \dot{q}_f}{2N_c} + \frac{\dot{\bar{q}}_f^2\de \dot{q}_f}{8N_c^3} \right) 
-  \left(\frac{k\de \dot{q}_i}{2N_c} + \frac{\dot{\bar{q}}_i^2\de \dot{q}_i}{8N_c^3} \right)\right\}
\biggr] \, ,
\eeq
where 
\beq
\label{eq:BC_name}
\bar{q}_i = \bar{q}(t=0) \, , 
\hspace{5mm}
\bar{q}_f = \bar{q}(t=1) \, ,
\hspace{5mm}
\dot{\bar{q}}_i = \dot{\bar{q}}(t=0) \, ,
\hspace{5mm}
\dot{\bar{q}}_f = \dot{\bar{q}}(t=1) \, .
\eeq
Note that the boundary terms are function of $\bar{q}_{(i,f)}$ and 
$\dot{\bar{q}}_{(i,f)}$ which are on-shell quantities. These are 
different from $q_{(i,f)}$ and $\dot{q}_{(i,f)}$ which also 
includes boundary values of the off-shell trajectories.
The bulk action given in eq. (\ref{eq:Sact_frw_simp_inp})
is used to determine the momentum conjugate to the field $q(t)$
\beq
\label{eq:mom_conju_qq}
\pi = \frac{\pt {\cal L}}{\pt \dot{q}} = - \frac{3\dot{q}}{2N_c} \, .
\eeq
In terms of conjugate momentum the boundary terms can be 
expressed as follows
\beq
\label{eq:Sbd_mom}
S_{\rm bdy} = - \ep\biggl[
\left(\bar{q}_f \de \pi_f - \bar{q}_i \de \pi_i \right)
+4 \al \left\{ 
\left(k\de \pi_f + \frac{\bar{\pi}_f^2\de \pi_f}{9} \right) 
-  \left(k\de \pi_i + \frac{\bar{\pi}_i^2\de\pi_i}{9} \right)\right\}
\biggr] \, .
\eeq
For the consistent variational problem these terms either have to be 
appropriately canceled by supplementing the original action 
given in eq. (\ref{eq:Sact_frw_simp})
with suitable surface terms or they vanish identically for the 
choice of boundary conditions.  

In the following, we will always impose Dirichlet boundary condition
at $t=1$, while at $t=0$ we will either have Neumann or Robin boundary 
conditions (as imposing Dirichlet at $t=0$ leads to unsuppressed 
perturbations, so it won't be addressed 
\cite{Feldbrugge:2017fcc,Feldbrugge:2017kzv}). 
However, as the path integral in two cases (NBC and RBC) are related by
eq. (\ref{eq:Grbc_Gnbc}) therefore one can first compute the 
path integral in NBC case and use the relation in eq. (\ref{eq:Grbc_Gnbc})
to compute the results for the Robin BC.

\subsection{Neumann Boundary condition (NBC) at $t=0$}
\label{neumann}

Imposing Neumann boundary condition \cite{Krishnan:2016mcj,DiTucci:2019dji} 
at $t=0$ and a Dirichlet boundary condition at $t=1$ gives a consistent variational 
problem. This means that the initial momentum $\pi_i$ and final 
position $q_f$ of all the trajectories are fixed.
This will mean
\beq
\label{eq:neuMa_cond}
\pi_i \,\, \&  \,\,
q_f = {\rm fixed} 
\hspace{5mm} 
\Rightarrow 
\hspace{5mm}
\de \pi_i = 0  \hspace{3mm} \&  \hspace{3mm}
\de q_f = 0 \, .
\eeq
Such a boundary condition imposition also has an additional 
merit that they lead to a well-behaved path integral where 
the perturbations are suppressed 
\cite{DiTucci:2019bui,Narain:2021bff,DiTucci:2020weq, Lehners:2021jmv}.
For this case, the boundary terms given in eq. (\ref{eq:Sbd_mom}) 
arising during the variation of action will reduce to the following 
\beq
\label{eq:Sbd_mom_NBC}
\biggl. S_{\rm bdy} \biggr|_{\rm NBC} = - \ep\biggl[
\bar{q}_f \de \pi_f
+4 \al 
\left(k\de \pi_f + \frac{\bar{\pi}_f^2\de \pi_f}{9} \right) 
\biggr] \, .
\eeq
These residual boundary terms which don't vanish after imposing boundary 
condition need to be canceled by a suitable addition of surface terms
so that the variational problem is consistent. 
It is seen that if one adds the following terms at the boundary 
\beq
\label{eq:Sact_surf_NBC}
\biggl. S_{\rm surface} \biggr|_{\rm NBC}
= \frac{1}{2} \biggl[
-\frac{3q_f \dot{q}_f}{N_c} 
- 24 \al \left(
\frac{k\dot{q}_f}{2N_c} + \frac{\dot{q}_f^3}{24 N_c^3} 
\right) 
\biggr] 
= q_f \pi_f 
+ 4 \al \left(k\pi_f +\frac{\pi_f^3}{27} \right)  \, ,
\eeq
then its variation (as suggested in eq. (\ref{eq:qfluc}) 
will precisely cancel the terms given in 
eq. (\ref{eq:Sbd_mom_NBC}). 
The first term in eq. (\ref{eq:Sact_surf_NBC}) is a
Gibbon-Hawking-York (GHY)-term and the second one 
is a Chern-Simon like term at the final boundary.

The constants $c_{1,2}$ that appear in the solution to 
the equation of motion eq. (\ref{eq:qsol_gen}) can now 
be determined for the choice of boundary conditions.  
This will imply
\beq
\label{eq:qsol_nbc}
\bar{q}(t) = \frac{\Lam N_c^2}{3} (t^2-1) - \frac{2 N_c \pi_i}{3} (t-1) + q_f \, ,
\eeq
where `bar' over $q$ is added as it is the solution to equation of motion.
In this setting at $t=0$, we notice that the on-shell value of 
$q_i$ is given by
\beq
\label{eq:q0_nbc}
\bar{q}_i = q_f + \frac{2 N_c \pi_i}{3} - \frac{\Lam N_c^2}{3} \, .
\eeq
For off-shell trajectories, $q_i$ can be anything as that is not fixed at $t=0$. 
The surface terms obtained in eq. (\ref{eq:Sact_surf_NBC}) when added to 
the action in eq. (\ref{eq:Sact_frw_simp_inp}) leads to the 
full action of the system. This is given by
\beq
\label{eq:Sact_frw_simp_nbc}
S_{\rm tot}[q,N_c]
=\frac{1}{2} \int_0^1 {\rm d}t \biggl[
(6 k - 2\Lam q) N_c - \frac{3 \dot{q}^2}{2N_c} \biggr] 
+ q_i \pi_i
+ 4 \al \left(k\pi_i +\frac{\pi_i^3}{27} \right)
\, ,
\eeq
Furthermore, on substituting the solution to the equation 
of motion eq. (\ref{eq:qsol_nbc}) and using eq. (\ref{eq:q0_nbc}) 
in the above, we will obtain 
the on-shell action which is given by
\beq
\label{eq:stot_onsh_nbc}
S_{\rm tot}^{\rm on-shell}[\bar{q}, N_c] = \frac{\Lam^2}{9} N_c^3 
- \frac{\Lam \pi_i}{3} N_c^2 
+ \left(3k - \Lam q_f + \frac{\pi_i^2}{3} \right) N_c
+ q_f \pi_i + 4\al \left(k\pi_i + \frac{\pi_i^3}{27} \right) \, .
\eeq
This is also the action for the lapse $N_c$. 
Compare this action with the action computed when 
Dirichlet boundary condition is imposed at $t=0$
\cite{Feldbrugge:2017kzv,Feldbrugge:2017fcc,Feldbrugge:2017mbc,
DiTucci:2019bui,Narain:2021bff,
Lehners:2021jmv, DiTucci:2020weq}
and notice the lack of singularity at $N_c=0$. 
The lack of singularity at $N_c=0$ can be physically 
understood as in the NBC path integral we sum over 
all possible transitions from $q_i$ (fixed $\pi_i$) 
to fixed $q_f$. This will also include the transition 
from $q_f$ to $q_f$ which can occur instantaneously {\it i.e.} with $N_c=0$.
Thereby implying that there is nothing singular happening at $N_c=0$.

\subsection{Robin Boundary condition (RBC) at $t=0$}
\label{robin_surf}

We now consider the situation when we have Robin boundary condition at 
$t=0$ and Dirichlet boundary condition imposed at $t=1$. 
An example of imposing Robin BC is fixing the Hubble parameter at a given time
in the cosmological evolution.
This boundary value 
problem poses a consistent variational problem. In this case, we fix a 
linear combination of the initial scale factor $q_i$ and the corresponding initial
conjugate momentum $\pi_i$, while the 
final scale factor $q_f$ is fixed at $t=1$. This means 
\beq
\label{eq:robin_cond}
\pi_i + \bt q_i = P_i= {\rm fixed} \hspace{3mm} \&  \hspace{3mm}
q_f = {\rm fixed} 
\hspace{5mm} 
\Rightarrow 
\hspace{5mm}
\de P_i = 0  \hspace{3mm} \&  \hspace{3mm}
\de q_f = 0 \, .
\eeq
For these boundary conditions, it has also been noted that perturbations 
are suppressed and in path integral their effects are bounded 
\cite{DiTucci:2019bui,Narain:2021bff,DiTucci:2020weq, Lehners:2021jmv}.
For the boundary conditions mentioned in eq. (\ref{eq:robin_cond}), 
the boundary terms generated during the variation of action 
given in eq. (\ref{eq:Sbd_mom}) reduces to the following 
\beq
\label{eq:Sbd_mom_RBC}
\biggl. S_{\rm bdy} \biggr|_{\rm RBC} = - \ep\biggl[
\bar{q}_f \de \pi_f + \bt \bar{q}_i \de q_i 
+4 \al \left\{ 
k\de \pi_f + \frac{\bar{\pi}_f^2\de \pi_f}{9} 
+ \bt\left(k\de q_i + \frac{(\bar{P}_i- \bt \bar{q}_i)^2\de q_i}{9} \right)\right\}
\biggr] \, .
\eeq
For the consistent variational problem, these terms need to be 
removed by supplementing the original gravity action with suitable 
surface terms. The additional suitable surface terms when varied will 
precisely cancel the above terms leading to a consistent variational 
problem. It is seen that if one adds the following terms at the boundary 
\bea
\label{eq:Sact_surf_RBC}
\biggl. S_{\rm surface} \biggr|_{\rm RBC} 
&=& q_f \pi_f 
+ 4 \al \left(k\pi_f +\frac{\pi_f^3}{27} \right) 
+ \frac{\bt}{2} (q_f^2 + q_i^2)
\notag \\
&&
+ 4 \al \bt \biggl[
k q_i + \frac{q_i}{9} \left(
P_i^2 - \bt P_i q_i + \frac{\bt^2 q_i^2}{3} 
\right)
\biggr]\, ,
\eea
Here the first two terms are the same as for the NBC case, while the rest of the terms
arise due to the imposition of RBC at $t=0$. In the limit of $\bt\to0$
these terms will vanish and we get back the surface terms for the NBC
which gives us a check of consistency. The variation of term $\bt q_f^2/2$ 
is although zero (due to imposition of DBC  at $t=1$), 
however its addition facilitates comparison with the 
results given in section \ref{rbc_free}. 
The surface term $\bt q_i^2/2$ 
in eq. (\ref {eq:Sact_surf_RBC}) comes due to the
RBC at $t=0$ from the Einstein-Hilbert part of the gravitational action
and was known from before \cite{DiTucci:2019dji},
The term proportional to $\al \bt$ is new and corresponds to surface terms 
that need to be added for the Gauss-Bonnet gravity part. 

The surface terms obtained in eq. (\ref{eq:Sact_surf_RBC}) when added to 
the action in eq. (\ref{eq:Sact_frw_simp_inp}) leads to the 
full action of the system. This is given by
\beq
\label{eq:Sact_frw_simp_RBC}
S_{\rm tot}[q,N_c]
=\frac{1}{2} \int_0^1 {\rm d}t \biggl[
(6 k - 2\Lam q) N_c - \frac{3 \dot{q}^2}{2N_c} \biggr] 
+ q_i P_i + \frac{\bt}{2} (q_f^2 - q_i^2)
+ 4 \al \left(k P_i +\frac{P_i^3}{27} \right)
\, .
\eeq
Comparing this action with the action for the NBC case given in eq. (\ref{eq:Sact_frw_simp_nbc}), we get
in the limit $\bt\to0$
\beq
\label{eq:btLimit_Sact}
\biggl. S_{\rm tot} \biggr|_{\rm RBC} 
\xRightarrow{\bt \to 0}
\biggl. S_{\rm tot} \biggr|_{\rm NBC} \, ,
\eeq
as $P_i \to \pi_i$. 
It should be mentioned that these boundary 
terms (which are called canonical boundary terms)
don't have a covariant expression 
(at least so far, it is not known) \cite{DiTucci:2019bui, DiTucci:2019dji}.
As gravity is a covariant theory, one would 
expect the boundary action to be correspondingly covariant. Indeed there exist 
other covariant boundary terms in 
literature satisfying the variational problem, which have been studied in 
\cite{Deruelle_2010, Krishnan:2017bte, Matsui:2023hei}.
These have been investigated in the mini-superspace 
path integral under certain situations in \cite{DiTucci:2019bui}.
However, the canonical terms when viewed from the 
quantum mechanical perspective 
\cite{DiTucci:2019dji} have an interpretation of a
complex coherent state, which becomes a plane 
wave in the limit $\beta \to 0$. 
Do note that such boundary terms 
also arise when one studies the one-dimensional 
quantum mechanical problem of a particle in an arbitrary potential 
with RBC as shown in section \ref{rbc_free}.
We will discuss more on the choice of $\beta$
and the usefulness of these boundary terms 
in section \ref{inter}. 

The constants $c_{1,2}$ that appear in the solution to 
the equation of motion given in eq. (\ref{eq:qsol_gen}) can now 
be determined for the case of RBC.
This will imply
\beq
\label{eq:qsol_RBC}
\bar{q}(t) = \frac{\Lam N_c^2}{3} t^2 
+ \frac{P_i}{\bt}
+ \left(1 + \frac{3}{2 \bt N_c} \right)^{-1}
 \left(t + \frac{3}{2 \bt N_c} \right)
 \left(q_f - \frac{P_i}{\bt} - \frac{\Lam N_c^2}{3} \right) \, ,
\eeq
where `bar' over $q$ indicates the solution of the equation of motion. 
Setting $t=0$ in this gives the on-shell value of the $q_i$ 
which is given by
\beq
\label{eq:q0_rbc}
\bar{q}_i =\frac{P_i}{\bt}
+\left(\frac{3}{3+2 \bt N_c} \right)
\left(q_f - \frac{P_i}{\bt} - \frac{\Lam N_c^2}{3} \right) \,.
\eeq
 Off-shell $q_i$ could be anything as it is not fixed by the boundary 
condition imposed in RBC. The on-shell action can be computed
by substituting the solution of the equation of motion and $\bar{q}_i$ 
into the action for RBC given in eq. (\ref{eq:Sact_frw_simp_RBC}). 
This is given by
\bea
\label{eq:stot_onsh_rbc}
S_{\rm tot}^{\rm on-shell}[\bar{q}, N_c] &=& \frac{1}{18(3+ 2 N_c \bt)} \biggl[
\bt \Lam^2 N_c^4 + 6 \Lam^2 N_c^3 +
N_c^2 \{108 \bt  k-18 \Lam  (P_i+\bt q_f)\}
\notag \\
&&
+18 N_c \left\{9
k+P_i^2+ q_f \left(\bt^2 q_f-3 \Lambda \right)\right\}
+ 54 P_i q_f
\biggr] 
+ 4 \al \left(k P_i +\frac{P_i^3}{27} \right) .
\eea
This is the action for the lapse $N_c$ in the case of RBC at $t=0$
and DBC at $t=1$.
In the limit $\bt\to0$, this reduces to the NBC on-shell action
given in eq. (\ref{eq:stot_onsh_nbc}).\\

\section{Transition Amplitude}
\label{trans}

We now move forward to compute the transition amplitude from 
one $3$-geometry to another (for both NBC and RBC case). 
The quantity that is of relevance here in the mini-superspace approximation
is as follows (see \cite{Halliwell:1988ik,Feldbrugge:2017kzv} for the Euclidean 
gravitational path integral in mini-superspace approximation)
\beq
\label{eq:Gamp}
G[ {\rm Bd}_f, {\rm Bd}_i]
= \int_{0^+}^{\infty} {\rm d} N_c  
\int_{{\rm Bd}_i}^{ {\rm Bd}_f} {\cal D} q(t) \,\, 
\exp \left(\frac{i}{\hbar} S_{\rm tot}[q, N_c] \right)
\, , 
\eeq
where ${\rm Bd}_i$ and  ${\rm Bd}_f$ are the initial and final 
boundary configurations respectively, and
$S_{\rm tot}$ refers to the mini-superspace action, which 
for the NBC case is given by eq. (\ref{eq:Sact_frw_simp_nbc}). 
We will first compute the expression for $G_{\rm NBC}[{\rm Bd}_f, {\rm Bd}_i]$
using eq. (\ref{eq:Gamp}) and (\ref{Gnbc_v}). We will then use 
the relation given in eq. (\ref{eq:Grbc_Gnbc}) to compute the 
expression for $G_{\rm RBC}[{\rm Bd}_f, {\rm Bd}_i]$.

\subsection{NBC at $t=0$}
\label{NBCt0}

To compute the expression for the transition amplitude in the case of 
Neumann boundary condition we make use of the results  
given in. eq. (\ref{Gnbc_v}). However, to make this 
connection we first compare the one-particle action written 
in eq. (\ref{act1part}) with the mini-superspace action for the 
NBC case given in eq. (\ref{eq:Sact_frw_simp_nbc}). 
This shows that the following substitution needs to be made 
\begin{gather}
\label{eq:subsNBC}
m \to -\frac{3}{2 N_c} \, , 
\hspace{1cm}
V(q) = \lam q \to \Lam N_c q 
\hspace{3mm} \Rightarrow \hspace{3mm} \lam \to \Lam N_c \, ,
\hspace{1cm}
p_i \to \pi_i \, .
\end{gather}
At this point, we are interested in computing 
\bea
\label{eq:NBC_qint}
\int_{{\rm Bd}_i}^{{\rm Bd}_f} {\cal D} q(t) \,\, 
&&
\exp \left(\frac{i}{\hbar} S_{\rm tot} [q, N_c]\right)
= \exp\left[\frac{i}{\hbar} \left\{3k N_c  
+ 4 \al \left(k \pi_i + \frac{\pi_i^3}{27}\right) \right\}\right]
\notag \\
&& \times \int_{{\rm Bd}_i}^{{\rm Bd}_f} {\cal D} q(t)
\exp\left[
\frac{i}{2\hbar} \int_0^1 {\rm d}t \biggl\{
- 2\Lam q N_c - \frac{3 \dot{q}^2}{2N_c} \biggr\}
+ \frac{i}{\hbar}q_i \pi_i
\right] \, .
\eea
This second line can be computed by exploiting the results given in 
eq. (\ref{Gnbc_v}) and making use of the substitutions mentioned 
in eq. (\ref{eq:subsNBC}). Note that the presence of term $q_i \pi_i$
with an integration over the initial $q_i$ leads to Fourier transform 
as expected in the NBC case. 
This gives the following expression for the 
Neumann boundary condition
\bea
\label{eq:gbar_NBC_ms}
&&
\int_{{\rm Bd}_i}^{{\rm Bd}_f} {\cal D} q(t) 
\exp \left(\frac{i}{\hbar} S_{\rm tot}[q, N_c] \right)
= \exp\left(\frac{i}{\hbar} S_{\rm tot}^{\rm on-shell}[\bar{q}, N_c] \right) 
\notag \\
&=& \exp\left[\frac{i}{\hbar} \left\{3k N_c  
+ 4 \al \left(k \pi_i + \frac{\pi_i^3}{27}\right) \right\}\right]
\bar{G}_{\rm NBC}[q_f, t=1; \pi_i, t=0] \, ,
\eea
where $S_{\rm tot}^{\rm on-shell}[\bar{q}, N_c]$ is given in eq. (\ref{eq:stot_onsh_nbc})
and $\bar{G}_{\rm NBC}[q_f, t=1; \pi_i, t=0]$ is. given by
\beq
\label{eq:GbarNBC_ms}
\bar{G}_{\rm NBC}[q_f, t=1; \pi_i, t=0]
= \exp \biggl[
\frac{i}{\hbar}\biggl\{
\frac{\Lam^2 N_c^3}{9} - \frac{\Lam \pi_i N_c^2}{3} 
+ \left(\frac{\pi_i^2}{3} - \Lam q_f\right) N_c + \pi_i q_f
\biggr\}
\biggr] \, .
\eeq
Notice that eq. (\ref{eq:gbar_NBC_ms}) this differs from 
the result obtained in \cite{Narain:2022msz} by a 
numerical prefactor. This difference in normalization is attributed to 
different styles of doing 
computation of path integral involving zeta-functions. The process of computing 
path integral via time-slicing is more reliable and correctly fixes the normalization.

Note $S_{\rm tot}^{\rm on-shell}[\bar{q}, N_c]$ doesn't become singular 
at $N_c=0$ and computation of the path integral over $q(t)$ doesn't 
give rise to additional functional dependence on $N_c$ which can be singular.
This allows us to extend the limits of $N_c$-integration all the way upto $-\infty$. 
This means that the transition amplitude is given by
\beq
\label{eq:Gab_afterQ}
G_{\rm NBC}[{\rm Bd}_f, {\rm Bd}_i]
= \frac{1}{2} \int_{-\infty}^\infty {\rm d} N_c \,\, 
\exp \left(\frac{i}{\hbar} S_{\rm tot}^{\rm on-shell}[\bar{q}, N_c] \right) \, ,
\eeq
where $S^{\rm on-shell}_{\rm tot}[\bar{q}, N_c]$ is given in eq. (\ref{eq:stot_onsh_nbc}).

To deal with the lapse integration in the NBC case we first make a change of variables.
This is done by shifting the lapse $N_c$ by a constant 
\beq
\label{eq:NcTONb}
N_c = \bar{N} + \frac{\pi_i}{\Lam} \, 
\hspace{5mm}
\Rightarrow
\hspace{5mm}
{\rm d}N_c \hspace{3mm} \to \hspace{3mm} {\rm d} \bar{N} \, .
\eeq
This change of variable correspondingly implies that the action for the lapse 
$S^{\rm on-shell}_{\rm tot}[\bar{q}, N_c]$ becomes the following 
\beq
\label{eq:sact_NcTONb}
S_{\rm tot}^{\rm on-shell}[\bar{q}, \bar{N}]
= \frac{\Lam^2}{9} \bar{N}^3 + (3k - \Lam q_f) \bar{N}
+ \left(\frac{3}{\Lam} + 4\al \right) \left(k \pi_i + \frac{\pi_i^3}{27} \right) \, .
\eeq
An interesting outcome of this is that after the change of variables the $\pi_i$
(initial momentum) dependence only appears in the constant term. 
After the change of variables, 
the transition amplitude is given by
\bea
\label{eq:Gab_Nb}
G[{\rm Bd}_f, {\rm Bd}_i]
= &&
\frac{1}{2} \exp \left[\frac{i}{\hbar} 
\left(\frac{3}{\Lam} + 4\al \right) \left(k \pi_i + \frac{\pi_i^3}{27} \right) \right]
\notag \\
&& \times
\int_{-\infty}^\infty {\rm d} \bar{N} \,\, 
\exp \left[\frac{i}{\hbar} \left\{\frac{\Lam^2}{9} \bar{N}^3 + (3k - \Lam q_f) \bar{N} \right\} \right] 
= \Psi_1(\pi_i) \Psi_2(q_f)
\, ,
\eea
where 
\bea
\label{eq:psi1}
&&
\Psi_1(\pi_i) = \frac{1}{2} \exp \left[\frac{i}{\hbar} 
\left(\frac{3}{\Lam} + 4\al \right) \left(k \pi_i + \frac{\pi_i^3}{27} \right) \right] \, ,
\\
\label{eq:psi2}
&&
\Psi_2(q_f) = \int_{-\infty}^\infty {\rm d} \bar{N} \,\, 
\exp \left[\frac{i}{\hbar} \left\{\frac{\Lam^2}{9} \bar{N}^3 + (3k - \Lam q_f) \bar{N} \right\} \right] \, .
\eea
The transition amplitude for the Neumann BC at $t=0$ and Dirichlet BC at the 
$t=1$ is a product of two parts: $\Psi_1(\pi_i) \Psi_2(q_f)$.
$\Psi_1(\pi_i)$ is entirely dependent on initial momentum $\pi_i$ 
and other $\Psi_2(q_f)$ is function of $q_f$ is related to the 
final size of the Universe. The dependence on two boundaries 
gets separated, a factorization also noticed in
\cite{Lehners:2021jmv} (and also in \cite{Narain:2022msz}) 
where the authors studied the 
Wheeler-DeWitt (WdW) equation in mini-superspace 
approximation of Einstein-Hilbert gravity.

\subsection{RBC at $t=0$}
\label{RBCt0}

We now come to the task of computing eq. (\ref{eq:Gamp})
for the case of RBC at $t=0$, where $S_{\rm tot}[q, N_c]$ for the RBC case 
is given in eq. (\ref{eq:Sact_frw_simp_RBC}). This mini-superspace 
RBC action can be compared with the quantum mechanical 
RBC problem discussed in subsection \ref{rbc_free}. This comparison 
leads to the substitution mentioned in eq. (\ref{eq:subsNBC}).
At this point, we are interested in computing 
\bea
\label{eq:RBC_qint}
\int_{{\rm Bd}_i}^{{\rm Bd}_f} 
&&
{\cal D} q(t) \,\, 
\exp \left(\frac{i}{\hbar} S_{\rm tot}[q, N_c] \right)
= \exp\left[\frac{i}{\hbar} \left\{3k N_c  
+ 4 \al \left(k P_i + \frac{P_i^3}{27}\right) \right\}\right]
\notag \\
&& \times \int_{{\rm Bd}_i}^{{\rm Bd}_f} {\cal D} q(t)
\exp\left[
\frac{i}{2\hbar} \int_0^1 {\rm d}t \biggl\{
- 2\Lam q N_c - \frac{3 \dot{q}^2}{2N_c} \biggr\}
+ \frac{i}{\hbar}q_i P_i + \frac{\bt}{2\hbar}(q_f^2 - q_i^2)
\right] 
\notag \\
&&
= \exp\left[\frac{i}{\hbar} \left\{3k N_c  
+ 4 \al \left(k P_i + \frac{P_i^3}{27}\right) \right\}\right] 
\bar{G}_{\rm RBC}[q_f, t=1; P_i, t=0]
\, ,
\eea
where
\bea
\label{eq:gbar_rbc_ms_qint}
\bar{G}_{\rm RBC}[q_f, t=1; P_i, t=0]
= && {\cal N} \int_{{\rm Bd}_i}^{{\rm Bd}_f} {\cal D} q(t)
\exp\biggl[
\frac{i}{2\hbar} \int_0^1 {\rm d}t \biggl\{
- 2\Lam q N_c - \frac{3 \dot{q}^2}{2N_c} \biggr\}
\notag \\
&&
+ \frac{i}{\hbar}q_i P_i + \frac{\bt (q_f^2-q_i^2)}{2\hbar}
\biggr] \,.
\eea
It is noted that the expression for $\bar{G}_{\rm RBC}[q_f, t=1; P_i, t=0]$
is exactly referring to the RBC path integral in terms of DBC path integral
with the substitution given in eq. (\ref{eq:subsNBC}).
Except that ${\cal N}$ has been added to get rid of 
$e^{i\bt q_f^2/2\hbar}$ due to requirement of classicality, where 
in a WKB sense, the amplitude is expected to become more classical 
as the Universe expands \cite{Feldbrugge:2017kzv, DiTucci:2019bui}.
However, using eq. (\ref{eq:Grbc_Gnbc}) one can express
$\bar{G}_{\rm RBC}[q_f, t=1; P_i, t=0]$ in terms of 
$\bar{G}_{\rm NBC}$ via the integral transform.
This means we have 
\bea
\label{eq:GbarRBC_nbc_ms}
\int_{{\rm Bd}_i}^{{\rm Bd}_f} 
&&
{\cal D} q(t) \,\, 
\exp \left(\frac{i}{\hbar} S_{\rm tot}[q, N_c] \right)
= \exp\left[\frac{i}{\hbar} \left\{3k N_c  
+ 4 \al \left(k P_i + \frac{P_i^3}{27}\right) \right\}\right]
\notag \\
&& \times \left(\frac{2\pi\hbar}{i \bt}\right)^{1/2}
\int_{-\infty}^{\infty} \frac{{\rm d} \tilde{p}}{2\pi\hbar}
e^{i (P_i - \tilde{p})^2/2 \hbar \bt}
\bar{G}_{\rm NBC}[q_f, t=1; \tilde{p}, t=0] \, .
\eea
The full path integral in the RBC case also involves 
integration over lapse $N_c$ with limits $(0,\infty)$.
This means 
\bea
\label{eq:Gamp_rbc_ms_exp}
G[{\rm Bd}_f, {\rm Bd}_i]&&
= \left(\frac{2\pi\hbar}{i \bt}\right)^{1/2} 
\exp\left[  
\frac{4 i \al}{\hbar}\left(k P_i + \frac{P_i^3}{27}\right) \right]
\int_{-\infty}^{\infty} \frac{{\rm d} \tilde{p}}{2\pi\hbar}
e^{i (P_i - \tilde{p})^2/2 \hbar \bt}
\notag \\
&&
\times
\int_{0^+}^{\infty} {\rm d} N_c 
\exp\left(\frac{3 k N_ci}{\hbar}\right)
\bar{G}_{\rm NBC}[q_f, t=1; \tilde{p}, t=0] \, .
\eea
As the $N_c$ integrand is not singular at $N_c=0$ so one 
can extend the integration limit all the way up to $-\infty$. 
The $N_c$-integration thereafter becomes similar to the integral 
studied in the case of NBC except the lack of Gauss-Bonnet term 
dependent on $\alpha$. This means one can write 
\beq
\label{eq:Gbar_nbc_EH_part}
\int_{0^+}^{\infty} {\rm d} N_c 
\exp\left(\frac{3 k N_ci}{\hbar}\right)
\bar{G}_{\rm NBC}[q_f, t=1; \tilde{p}, t=0]
= \biggl. \Psi_1(\tilde{p}) \biggr|_{\al =0} \times \Psi_2(q_f) \, ,
\eeq
where $\Psi_1$ and $\Psi_2$ are given in 
eq. (\ref{eq:psi1}) and (\ref{eq:psi2}) respectively.
Here again, as the integrand doesn't have $N_c=0$ singularity,
the integral can be extended all to way to $-\infty$ and including an 
extra $1/2$ factor which is absorbed in the definition of $\Psi_1$.
The leftover integral is the integral over $\tilde{p}$. This integral 
can be cast into a more familiar form by re-definition of $\tilde{p}$ as
\beq
\label{eq:pt_to_pb}
\tilde{p} \to \bar{p} - \frac{3 \Lam}{2\bt} \, .
\eeq
Such a transformation allows us to rewrite the $\tilde{p}$ integral 
as an Airy integral. This is given by
\beq
\label{eq:pt_int_airy_rbc}
\int_{-\infty}^{\infty} \frac{{\rm d} \tilde{p}}{2\pi\hbar}
e^{i (P_i - \tilde{p})^2/2 \hbar \bt} \biggl. \Psi_1(\tilde{p}) \biggr|_{\al =0}
= \frac{1}{2} 
\exp\biggl[
\frac{i\bigl(
-18 k \bt^2 + 2 P_i^2 \bt^2 
+ 6 P_i \bt \Lam + 3 \Lam^2
\bigr)}{4 \hbar \bt^3} 
\biggr]
\Phi(P_i, \bt)\, .
\eeq
where 
\beq
\label{eq:Phi_Pi_beta}
\Phi(P_i, \bt)
= \int_{-\infty}^{\infty} \frac{{\rm d} \bar{p}}{2\pi\hbar}
\exp\biggl[
\frac{i}{\hbar} \biggl\{
\frac{\bar{p}^3}{9\Lam} 
+ \left(\frac{3k}{\Lam} - \frac{P_i}{\bt} - \frac{3\Lam}{4\bt^2} \right) \bar{p}
\biggr\}
\biggr] \, .
\eeq
Putting all the pieces together give
\bea
\label{eq:Gamp_rbc_int_exact}
G[ {\rm Bd}_f, {\rm Bd}_i]
&&
= \frac{1}{2} \left(\frac{2\pi\hbar}{i \bt}\right)^{1/2}
\exp\left[  
\frac{4 i \al}{\hbar}\left(k P_i + \frac{P_i^3}{27}\right) \right]
\exp\biggl[
\frac{i\bigl(
-18 k \bt^2 + 2 P_i^2 \bt^2 
+ 6 P_i \bt \Lam + 3 \Lam^2
\bigr)}{4 \hbar \bt^3} 
\biggr]
\notag \\
&& 
\times 
\Phi(P_i, \bt)  \,\,
\Psi_2(q_f) \, .
\eea
It is noticed that for the RBC case too the transition amplitude gets 
factorized in two parts: one dependent on the final boundary 
and one dependent on the initial boundary.
In the next sub-section, we will compute these integrals in terms of Airy functions. 
It is expected that the RBC transition amplitude will be a 
product of two Airy functions along with exponential 
prefactors.

In order to see the $\bt\to0$ limit clearly it is more cleaner to write the 
$\bt$-dependent terms in eq. (\ref{eq:GbarRBC_nbc_ms}) as follows
\beq
\label{eq:beta_change}
\left(\frac{2\pi\hbar}{i \bt}\right)^{1/2} e^{i (P_i - \tilde{p})^2/2 \hbar \bt}
= \int_{-\infty}^\infty {\rm d} \xi \exp\left[
-\frac{i \bt}{2\hbar} \xi^2 + \frac{i}{\hbar} (P_i -  \tilde{p}) \xi
\right] \, .
\eeq
In this way, the limit $\bt\to0$ gives a $\de$-function $\de(P_i - \tilde{p})$.
This limit is more harder to see from the end result of transition amplitude 
as one has to work with asymptotic forms of the Airy-functions.

\subsection{Airy function}
\label{airy}

In this sub-section, we will compute the integrals 
$\Psi_2(q_f)$ and $\Phi(P_i, \bt)$ given in eq. (\ref{eq:psi2})
and (\ref{eq:Phi_Pi_beta}) respectively. It should be noted that these functions 
can be identified with the \textit{Airy}-integrals. 
These integrals are sensitive to the contour of integration. 
In the case of Airy-integrals, the regions of convergence are 
within the following phase angles
$\ta \equiv \arg(\bar{N})$: $0\leq \ta \leq \pi/3$ (region $1$),
$2\pi/3 \leq \ta \leq \pi$ (region $0$), and 
$4\pi/3 \leq \ta \leq 5\pi/3$ (region 2). One can define the 
following contours: ${\cal C}_0$ the contour 
running from region $0$ to region $1$, 
${\cal C}_1$ the contour 
running from region $1$ to region $2$, and 
${\cal C}_2$ the contour 
running from region $2$ to region $0$.
By making use of the above contours of integration 
one can define the following integrals
\bea
\label{eq:airy_ai}
&&
Ai(z) = \frac{1}{2\pi} \int_{{\cal C}_0} {\rm d} x \exp
\left[i\left(\frac{x^3}{3} + z x \right) \right] \, , 
\\
\label{eq:airy_bi}
&&
Bi(z) = \frac{i}{2\pi} \int_{{\cal C}_2 - {\cal C}_1} {\rm d} x \exp
\left[i\left(\frac{x^3}{3} + z x \right) \right] \, .
\eea
$\Psi_2(q_f)$ can be computed as discussed in detail in the paper 
\cite{Narain:2022msz}. It is given by
\beq
\label{eq:Psi2_rbc}
\Psi_2(q_f) =  \sqrt{3}\left(\frac{3\hbar}{\Lam^2}\right)^{1/3}
Ai\left[
\left(\frac{\sqrt{3}}{\hbar \Lam} \right)^{2/3} \left(3k - \Lam q_f \right)
\right] \, ,
\eeq
The computation for $\Phi(P_i, \bt)$ can be done in an analogous manner.
It is given by,
\beq
\label{eq:Phi_pi_bt_exp}
\Phi(P_{i},\bt) = \left(\frac{3\Lam}{\hbar^2}\right)^{1/3} 
Ai \biggl[
\left(\frac{3\Lam}{\hbar^2}\right)^{1/3}
\left(\frac{3k}{\Lam}-\frac{P_{i}}{\bt}-\frac{3\Lam}{4\bt^2}\right)
\biggr] \, .
\eeq
Putting these expressions together gives the exact expression 
for the transition amplitude for the NBC and RBC case.
These are given by,
\bea
\label{eq:Gbd0bd1_full_nbc}
G_{\rm NBC}[{\rm Bd}_i, {\rm Bd}_f]
= &&
\frac{\sqrt{3}}{2} \left(\frac{3\hbar}{\Lam^2}\right)^{\frac{1}{3}}
\exp \left[\frac{i}{\hbar} 
\left(\frac{3}{\Lam} + 4\al \right) \left(k \pi_i + \frac{\pi_i^3}{27} \right) \right]
\notag \\
&&
\times
Ai\left[
\left(\frac{3}{\hbar^2 \Lam^2} \right)^{\frac{1}{3}} \left(3k - \Lam q_f \right)
\right] \, ,
\\
\label{eq:Gbd0bd1_full_rbc}
G_{\rm RBC}[{\rm Bd}_i, {\rm Bd}_f]
= &&
\sqrt{\frac{3\pi\hbar}{2i \bt}}
\exp\left[  
\frac{4 i \al}{\hbar}\left(k P_i + \frac{P_i^3}{27}\right) \right]
\exp\biggl[
\frac{i\bigl(
-18 k \bt^2 + 2 P_i^2 \bt^2 
+ 6 P_i \bt \Lam + 3 \Lam^2
\bigr)}{4 \hbar \bt^3} 
\biggr] 
\notag \\
&&
\hspace{-15mm}
\times
\left(\frac{9}{\Lam\hbar}\right)^{\frac{1}{3}}
Ai\left[
\left(\frac{3}{\hbar^2 \Lam^2} \right)^{\frac{1}{3}} \left(3k - \Lam q_f \right)
\right] 
Ai \biggl[
\left(\frac{3\Lam}{\hbar^2}\right)^{\frac{1}{3}}
\left(\frac{3k}{\Lam}-\frac{P_{i}}{\bt}-\frac{3\Lam}{4\bt^2}\right)
\biggr] \, .
\eea
This is an exact result for the case when NBC and RBC are imposed 
at initial time $t=0$ respectively. Notice also the correction coming 
from the Gauss-Bonnet sector of the gravitational action which appears 
as an exponential prefactor. It is also crucial to notice that this 
GB correction doesn't appear in the Airy functions which only depend
on the boundary conditions. 

\section{$\hbar\to0$ limit}
\label{hbar0}

In this section, we will study the $\hbar\to0$ limit of the exact wave-function 
computed in the previous section. It is relevant to look at this as in the 
$\hbar\to0$ limit the exact computation should reproduce known features 
and should agree with the results coming from saddle-point approximation 
which will be discussed in more detail in section \ref{sad_pot_approx}. 
The $\hbar\to0$ limit also highlights the configuration space which gives 
the dominant contribution in the path integral. This ultimately translates 
into preferable values for the initial boundary parameters. One can then 
do the saddle point analysis for these values of initial parameters 
which gives the dominant contribution in the path integral.

We will start by analyzing the nature of 
the contribution coming from the Gauss-Bonnet sector and the 
constraints that come from it in the limit $\hbar\to0$.

\subsection{Gauss-Bonnet contribution}
\label{GBcont}

\begin{figure}[h]
\centerline{
\vspace{0pt}
\centering
\includegraphics[width=10cm]{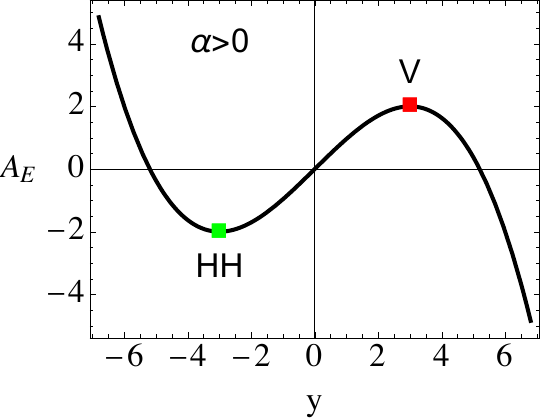}
}
\caption[]{
Plot of Euclidean action $A_E$ given in eq. (\ref{eq:Aeuc}) vs $y$. 
The two saddle points are depicted by green and red squares. 
For $\al>0$: green one corresponds to $P_i = -3 i \sqrt{k}$ (stable) 
while the red one corresponds to $P_i = 3 i \sqrt{k}$ (unstable). 
Interestingly, $P=-3i \sqrt{k}$ is also the value for which geometry 
become regularised at $t=0$ and perturbations are well-behaved.
}
\label{fig:aept}
\end{figure}
%

The first thing we will focus on is the exponential prefactor that 
includes the effects coming from the Gauss-Bonnet sector. This 
prefactor is the same in both NBC and RBC cases ($P_i \to \pi_i$ in 
limit $\bt\to0$). This prefactor is the following
\beq
\label{eq:expGBfac}
\exp\left[  
\frac{4 i \al}{\hbar}\left(k P_i + \frac{P_i^3}{27}\right) \right]
= e^{i A(P_i)/\hbar} \, ,
\eeq
where  $A(P_i)$ is given by
\beq
\label{eq:APi_act}
A(P_i) = 4 \al \left(k P_i + \frac{P_i^3}{27}\right) \, .
\eeq
In the limit $\hbar\to0$ the dominant contribution comes 
from configuration which extremizes $A(P_i)$, which
corresponds to the 
\beq
\label{eq:exptA_pi}
A'(P_i) =0 \hspace{3mm} \Rightarrow \hspace{3mm}
P_i = \pm 3 i \sqrt{k} \, .
\eeq
In the limit $\bt\to0$ this corresponds to $\pi_i = \pm 3 i \sqrt{k}$!
Note that $\pi_i = -3i \sqrt{k}$ also corresponds to the required value of 
the initial momentum for regular geometries with well-behaved 
perturbations in the case of Hartle-Hawking (HH) wave-function 
for the NBC \cite{Narain:2022msz} 
($3i \sqrt{k}$ corresponds to Vilenkin, where perturbations 
are unstable). 
These two signs of $P_i$ (or, $\pi_i$) correspond 
to two different orientations of wick rotations.

It should be emphasised that the $\hbar\to 0$ limit imposes 
classicalization when the quantum system starts behaving 
classically. This same end state where the system behaves 
like a classical system can also be achieved 
when $\hbar$ is not small, but the Gauss-Bonnet coupling is large.
In this situation, when the Gauss-Bonnet coupling $\al$ is large, 
the dominant contribution comes from the configuration whose $P_i$ lies around 
$P_i = \pm 3 i \sqrt{k}$. However, this situation is entirely quantum 
in nature, as $\hbar$ is not small while the system acquires 
the end state of ``classicality"  in the limit of large Gauss-Bonnet coupling. 
This can happen in the very early stages of the Universe.

Let us consider the rotated $P_i = i y$, which 
is like considering an Euclideanised version action $A$. 
Then we have the Euclidean action $A_E(y)$ given by
\beq
\label{eq:Aeuc}
A_E(y) = 4 \al \left(ky - \frac{y^3}{27} \right) \, ,
\hspace{5mm}
e^{i A(P_i)/\hbar}
\hspace{3mm} \Rightarrow \hspace{3mm}
e^{- A_E(y)/\hbar} \, .
\eeq
On extremization of Euclidean action $A_E(y)$ it is seen that 
for $\al>0$
the point $y=-3\sqrt{k}$ corresponds to a stable saddle point and leads to 
an exponential with a positive argument (Hartle-Hawking) while 
$y=3\sqrt{k}$ correspond to unstable saddle point leading to 
exponential with a negative argument (Vilenkin Tunneling). 
Coincidently, as was also mentioned in \cite{DiTucci:2019bui, Narain:2022msz}
$P_i=-3i \sqrt{k}$ also correspond to initial condition 
for which geometry becomes regularised at $t=0$ and
perturbations are well-behaved within a certain regime of $\beta$, as discussed in subsequent sections.

\subsection{Choice of $\bt$}
\label{btchoice}

The next question that one needs to address is that for the given 
$P_i=-3 i \sqrt{k}$ (as required by stability), what are the allowed possibilities for the $\bt$
in the case of RBC transition amplitude. This issue doesn't arise 
in the case of NBC as $\bt=0$ for the NBC transition amplitude.
To analyze the constraints (allowed values) that get imposed on $\bt$
from the $\hbar\to0$ limit, it is sufficient to study the nature 
of RBC wave-function at $t=0$ for the fixed allowed value 
of $P_i=-3i \sqrt{k}$.

We start by considering the $\hbar\to0$ limit of $\Phi(P_i, \bt)$
which for $P_i = -3i\sqrt{k}$ can be written as
\beq
\label{eq:Phi_pi_hbar}
\Phi(-3i \sqrt{k}, \bt)
= \left(3\Lam \hbar\right)^{1/3} 
Ai \biggl[
\left(\frac{9\sqrt{\Lam}}{\hbar}\right)^{2/3}
\left(\sqrt{\frac{k}{\Lam}} + \frac{i \sqrt{\Lam}}{2\bt} \right)^2
\biggr] \, .
\eeq
For imaginary $\bt$ the argument of Airy-function is always
positive for $\Lam>0$. This observation allows us to 
obtain easily the $\hbar\to0$ limit by exploiting the 
asymptotic structure of the Airy-functions with 
positive arguments
\beq
\label{eq:Phi_hbar_lim}
\biggl. \Phi\left(-3i \sqrt{k}, \bt \right) \biggr|_{\hbar\to0}
\sim 
\left(3\Lam \hbar\right)^{\frac{1}{3}} \exp\biggl[
- \frac{6\sqrt{\Lam}}{\hbar}
\left|\sqrt{\frac{k}{\Lam}} + \frac{i \sqrt{\Lam}}{2\bt} \right|^3
\biggr] 
= \left(3\Lam \hbar\right)^{\frac{1}{3}}  \exp\biggl[
-\frac{B_1(\bt)}{\hbar} 
\biggr]
\, ,
\eeq
where 
\beq
\label{eq:B1beta_func_def}
B_1(\bt)
= 6\sqrt{\Lam}\left|\sqrt{\frac{k}{\Lam}} + \frac{i \sqrt{\Lam}}{2\bt} \right|^3 \, ,
\eeq
and the $\sim$ sign implies that we are ignoring the numerical prefactor 
arising from the asymptotic form of Airy's function, which is not relevant for 
the following discussion. From the asymptotic structure, one could 
see that in limit $\hbar\to0$ the dominant contribution comes 
from those configurations for which $B_1^\prime(\bt)=0$. This means we have 
\beq
\label{eq:Bprime_bt}
B_1^\prime(\bt) = 18 \sqrt{\Lam}
\left(\sqrt{\frac{k}{\Lam}} + \frac{i \sqrt{\Lam}}{2\bt} \right)^2
\left(\frac{-i \sqrt{\Lam}}{2\bt^2}\right) = 0 \, .
\eeq
This means that the dominant contribution comes 
from configuration when $\bt$ lies around 
$\bt_{\rm dom} = -i \Lam /2\sqrt{k}$ (we don't 
consider the $\bt\to\infty$ case).

We next consider the exponential prefactor independent of $\al$
for $P_i=-3i\sqrt{k}$. This means that we have to investigate 
\beq
\label{eq:expfac_noa}
\exp\left[
\frac{
i \left(3\Lam^2 - 18 i \bt \Lam \sqrt{k} - 36 k \bt^2 \right)}
{4 \hbar \bt^3} 
\right]
= \exp\left(\frac{i B_2(\bt)}{\hbar} \right) \, ,
\eeq
where 
\beq
\label{eq:B2beta_func_def}
B_2(\bt)
=  \frac{\left(3\Lam^2 - 18 i \bt \Lam \sqrt{k} - 36 k \bt^2 \right)}
{4 \bt^3} \, .
\eeq
Once again in the $\hbar\to0$ limit the dominant 
contribution comes from those configurations for which $\bt$ 
lies around the extrema given by $B_2^\prime(\bt)=0$. This 
given $\bt_{\rm dom} = -i \Lam /2\sqrt{k}$ same as before.
This eventually give rise to two cases: $i\bt< i\bt_{\rm dom}$ 
and $i\bt>i\bt_{\rm dom}$ (note $\bt$ and $\bt_{\rm dom}$ are imaginary). 
To study these cases carefully let us write 
\beq
\label{eq:beta_x_dep}
\bt = -\frac{i \Lam x}{2\sqrt{k}} 
= \bt_{\rm dom} x
\hspace{5mm} {\rm where} \hspace{5mm} x\geq 0 \, .
\eeq
This gives
\beq
\label{eq:xformB1B2}
B_1(\bt) = 6\frac{k^{3/2}}{\Lam} \biggl| 1 - \frac{1}{x} \biggr|^3 \, ,
\hspace{10mm}
B_2(\bt) = -\frac{6 i k^{3/2} \left(3x^2 -3x +1 \right)}{x^3 \Lam} \, .
\eeq
At this point, it is worthwhile to note that according to the above analysis, purely (negative) imaginary initial momentum automatically enforces that the physically interesting range of $\beta$ has to lie also in the purely (negative) imaginary direction. Imaginary $\beta$ reminds the quantum nature of the initial condition. In the subsequent analysis, we will identify the specific sub-region that is physically relevant to our paper. Now, we consider cases when $x\leq 1$ and $x>1$, respectively. 

\subsubsection{$0\leq x \leq 1$}
\label{btlessBtdom}

In this case, we will have 
\beq
\label{eq:B1+B2_case1}
 \exp\biggl(
-\frac{ B_1(\bt)}{\hbar} 
\biggr)
\exp\left(\frac{i B_2(\bt)}{\hbar} \right)
= \exp\biggl[
\frac{6k^{3/2}}{\hbar \Lam}
\biggr] \, .
\eeq
This can be combined with the exponential prefactor coming from the 
Gauss-Bonnet sector of gravity ($\al$-dependent part) which for 
$P_i=-3 i \sqrt{k}$ is given by $\exp(8k^{3/2} \al/\hbar)$.
Combining the two exponential prefactors gives the following
\beq
\label{eq:HHrbcFactor}
\exp\left(\frac{6 k^{3/2}}{\hbar\bar{\Lam}}\right)
\hspace{3mm} \text{where,} \,\, 
\bar{\Lam}= \Lam/\left(1+ 4\al \Lam/3\right) \, .
\eeq
This is precisely the Hartle-Hawking state with positive weighting. 
We note that the Gauss-Bonnet modification doesn't prevent the 
no-boundary Universe solution to exist whose contribution 
appears as a multiplicative factor
(exponential weight). This higher-derivative gravity
correction (which, although is topological) 
further supports the findings in the 
paper \cite{Jonas:2020pos}, where the 
authors found the no-boundary Universe 
even after the inclusion of generic higher-derivative terms.
The topological nature of Gauss-Bonnet gravity, however 
allow us to go beyond the perturbative analysis 
done in \cite{Jonas:2020pos}.
It is worth emphasizing that the exponential prefactor is independent of $\beta$. 
The positive exponent says that lower $\Lam$ is more favorable, 
i.e., low values of potential are preferred. The analysis also includes the case 
$\bt=0$ i.e., Neumann BC. For a special value of $\al=- 3/4\Lam$, 
this HH-factor becomes unity.

\subsubsection{$x>1$}
\label{btmoreBtdom}

In this case, we will have 
\beq
\label{eq:B1+B2_case2}
 \exp\biggl(
-\frac{ B_1(\bt)}{\hbar} 
\biggr)
\exp\left(\frac{i B_2(\bt)}{\hbar} \right)
= \exp\biggl[
\frac{6k^{3/2}}{\hbar \Lam}
\left(
-1 + \frac{6}{x} - \frac{6}{x^2} + \frac{2}{x^3}
\right)
\biggr] \, .
\eeq
This exponential factor has to be multiplied by the 
contribution coming from the Gauss-Bonnet sector 
$e^{8 k^{3/2} \al/\hbar}$ to get the full overall 
exponential prefactor.
In the large $x\to\infty$ limit, we get inverse Hartle-Hawking. This configuration represents the tunneling geometry, and perturbations are unstable \cite{Feldbrugge:2017fcc}.
However, for other values of $x>1$, the nature of exponential 
depend on the behavior of the functional form of $x$. 
Let us call this function 
\beq
\label{eq:fx_beta_form}
f(x) = -1 + \frac{6}{x} - \frac{6}{x^2} + \frac{2}{x^3} \, .
\eeq
We notice that $f(1)=1$ and $f(\infty) = -1$. It should be mentioned 
that $f'(x) = - 6(x-1)^2/x^4 <0$ for all values $x$ except $x=1$ and $x=\infty$. 
This means $f(x)$ is a monotonically decreasing function of $x$ within the 
range between $1$ and $-1$. At some point 
\beq
\label{eq:x_x0_crossover}
x_0 = 2 + 2^{1/3} + 2^{2/3} \, ,
\eeq
the function $f(x)$ changes sign and becomes negative for $x>x_0$
while it remains positive for $1<x<x_0$ (although it is not equal to 
$1$, thereby implying that it doesn't have 
the actual Hartle-Hawking factor $6k^{3/2}/\hbar \bar{\Lam}$ as in eq. (\ref{eq:HHrbcFactor})).

\subsection{Interpretation}
\label{inter}

Before proceeding further, let us mention a 
possible interpretation of the Robin boundary condition in 
the context of no boundary proposal that is immediate 
from the above choice of $\beta$ (purely imaginary).
In this case, one can interpret 
the boundary terms as being complex coherent state {\it i.e.}, 
\begin{equation}
   \exp\left[i\left(-\frac{\beta q_i^2}{2}
   +P_iq_i\right)\right] \xrightarrow[]{\text{for,\,\,} 
   \beta=-i|\beta|}\exp\left[-\frac{|\beta| q_i^2}{2}+iP_iq_i\right] \, .
\end{equation}
This is a Gaussian state with an imaginary 
momentum $P_i$, peak at $q_i=0$ and 
uncertainty determined by $\beta (\Lambda)$ 
\cite{DiTucci:2019bui}. It describes a state with shared uncertainty 
between the scale factor and the momentum 
of the universe. In $\beta\to0$, this state 
becomes a plane wave, which is a momentum eigenstate. 
Taking this as an initial state for the Robin boundary 
condition \cite{DiTucci:2019bui, DiTucci:2019dji},
one can express the final state as a path-integral 
in the following manner
\begin{equation}
 \Psi[q_f,\beta,P_i] = \int dN 
 \mathcal{D}q \,dq_i e^{iS_{DD}[q_f,N,q]/\hbar} 
 \, \psi_0[P_i,\beta], \quad \psi_0[P_i,\beta] 
 \propto e^{i\left(- \frac{\beta q_i^2}{2}+P_iq_i\right)},
\end{equation}
where $S_{DD}$ is action with imposition of Dirichlet BC
at the two endpoints. 
In the limit $\lvert \beta \rvert \to \infty$, 
$e^{-i \beta q_i^2 /2} \to \delta(q_i)$. This is a 
sharp imposition of the boundary condition $q_i=0$,
which is the Dirichlet condition. The other
limit $\lvert \beta \rvert \to 0$ gives Neumann BC. 
Finite values of $\beta$ act as a kind of `regulator' 
embedding a regularized version of $\delta$-function 
in the path-integral. In a sense, RBC is a 
regularised DBC. 
This is also compatible with quantum uncertainty 
principle in the sense 
knowing the $q_i$ arbitrarily accurate 
would render the initial 
momentum completely undetermined. 
Also, quantum uncertainty doesn't allow
to determine the initial size with arbitrary accuracy \cite{DiTucci:2019dji}. 
In a way, studying Robin BC is the 
most appropriate scenario due to its compatibility 
with the quantum uncertainty principle and the 
regularized way of introducing Dirichlet BC 
overcoming technical complications associated with 
dealing with path integrals involving $\delta$-function. 

\section{Saddle-point approximation}
\label{sad_pot_approx}

The analysis of $\hbar\to0$ has shown us that certain initial configurations
are favorable and give the dominant contribution in the path integral.
The $\hbar\to0$ limit of the Gauss-Bonnet contribution shows 
that the dominant contribution comes 
from $P_i= \pm3 i \sqrt{k}$. Of which only the $P_i = -3 i \sqrt{k}$ 
correspond to a stable configuration. Coincidentally, this is also the 
same required value of $P_i$ which leads to stable well-behaved 
fluctuations in the Hartle-Hawking no-boundary proposal of the Universe
\cite{DiTucci:2019bui,Narain:2021bff, Lehners:2021jmv, 
DiTucci:2020weq, DiTucci:2019dji, Narain:2022msz}. 
In a sense, the Gauss-Bonnet presence favors the 
stable Hartle-Hawking no-boundary Universe. 

In this section, we will do the saddle point analysis of the gravitational 
path integral in the mini-superspace approximation. We will do this 
for the case of Robin boundary condition as the case of Neumann 
boundary condition has been already investigated in \cite{Narain:2022msz}.
However, the saddle-point study of the path integral will be done in a slightly 
different manner. The gravitational path integral in the mini-superspace 
approximation in the RBC case is given in 
eq. (\ref{eq:Gamp}) where the path integral over the $q(t)$ is given in 
eq. (\ref{eq:RBC_qint}). We use the methods developed in analyzing 
the quantum-mechanical RBC problem to convert eq. (\ref{eq:RBC_qint})
into eq. (\ref{eq:GbarRBC_nbc_ms}), which relates the RBC path integral 
of $q(t)$ to the NBC path integral of the same gravitational theory. 
Doing lapse $N_c$ integration over this gives the full transition amplitude,
relating the RBC transition amplitude with the NBC one as given in 
eq. (\ref{eq:Gamp_rbc_ms_exp}). This allows us to 
write the full RBC transition amplitude as a product 
of two integrals after we make use of eq. (\ref{eq:Gbar_nbc_EH_part}).
One is integral over $\tilde{p}$ and other integral over 
$\bar{N}$, where $\bar{N}$-integral is given by eq. (\ref{eq:psi2}).
Our strategy in doing the saddle-point analysis is to analyze 
each of these integrals (integral over $N_c$ and integral 
over $\tilde{p}$) separately. 

We first take a look at the integral over $N_c$ 
given in eq. (\ref{eq:Gbar_nbc_EH_part}). 
If one shifts $N_c$ as in eq. (\ref{eq:NcTONb}) then 
this integral becomes an integral over $\bar{N}$ and an 
exponential prefactor dependent on $\tilde{p}$. 
The $\bar{N}$-integral depends on $q_f$ written as 
$\Psi_2(q_f)$. The saddle point analysis of this 
has already been done in \cite{Narain:2022msz} and won't be 
repeated here. The results from the saddle analysis 
revealed an interesting feature. For all values of 
$q_f$ there are two saddle points, however, for 
$q_f<3k/\Lam$ only one saddle point which lies on the 
imaginary axis is relevant, while for $q_f>3k/\Lam$ there are two 
relevant saddle points lying on the real axis. This means 
that for $q_f<3k/\Lam$ Universe is Euclidean 
as can be seen from the exponential form of the transition amplitude
indicating imaginary `time', while for 
$q_f>3k/\Lam$ Universe is Lorentzian as is noticed from 
the oscillatory nature of transition amplitude indicating 
real time \cite{Narain:2022msz}.

Our task then shifts to analyzing the integral over 
$\tilde{p}$ given in eq.(\ref{eq:pt_int_airy_rbc}). 
\beq
\label{eq:Bint_def_Bact}
\int_{-\infty}^{\infty} \frac{{\rm d} \tilde{p}}{2\pi\hbar}
e^{i (P_i - \tilde{p})^2/2 \hbar \bt} \biggl. \Psi_1(\tilde{p}) \biggr|_{\al =0}
= \int_{-\infty}^{\infty} \frac{{\rm d} \tilde{p}}{2\pi\hbar}
e^{i B(\tilde{p})/\hbar}
=  \int_{-\infty}^{\infty} \frac{{\rm d} \tilde{p}}{2\pi\hbar}
e^{\{h(\tilde{p}) + i H(\tilde{p}) \}/\hbar} \, ,
\eeq
where
\beq
\label{eq:argExp_Pt_phi}
B(\tilde{p}) = \frac{(P_i - \tilde{p})^2}{2\bt} + \frac{3}{\Lam} 
\left(
k \tilde{p} + \frac{\tilde{p}^3}{27}
\right) \, .
\eeq
and $h(\tilde{p})$ is the corresponding Morse-function while the 
$H(\tilde{p})$ corresponds to real part of the $B$-function.
By a shift of variable as stated in eq. (\ref{eq:pt_to_pb}), 
the $\tilde{p}$-integral can be cast into an Airy integral along 
with an exponential prefactor, as discussed in section \ref{RBCt0}. 
Here, we will study integral 
in eq. (\ref{eq:Bint_def_Bact}) using 
Picard-Lefschetz methods (see \cite{Feldbrugge:2017kzv, Narain:2021bff, 
Witten:2010cx, Witten:2010zr, Basar:2013eka, Tanizaki:2014xba} 
for review on Picard-Lefschtez and analytic continuation). 

The saddle-points of $\tilde{p}$ can be obtained by computing 
the expression ${\rm d} B(\tilde{p})/{\rm d} \tilde{p}$. The saddle points 
are computed from the equation
\beq
\label{eq:sad_pt_B}
\frac{{\rm d} B(\tilde{p})}{{\rm d} \tilde{p}} = 
\frac{\tilde{p}^2}{3\Lam} + \frac{\tilde{p}}{\bt}
+ \left(\frac{3k}{\Lam} - \frac{P_i}{\bt} \right) =0 \, .
\eeq
This is a quadratic equation in $\tilde{p}$ resulting in two saddle points.
The discriminant $\D$ of the above quadratic equation is 
given by
\beq
\label{eq:disc_pt_Pi_bt_B}
\D = \frac{1}{\bt^2} - \frac{4k}{\Lam^2} + \frac{4 P_i}{3\bt \Lam} \, .
\eeq
For stable configuration referring to Hartle-Hawking no-boundary 
Universe ($P_i=-3 i \sqrt{k}$) and $\bt$ given by eq. (\ref{eq:beta_x_dep}),
the discriminant $\D$ becomes the following
\beq
\label{eq:Delta_x_nbp}
\biggl. \D \biggr|_{\rm Hartle-Hawking} = 
- \frac{4k(x-1)^2}{x^2 \Lam^2} \, .
\eeq
For all values of $x>0$, the discriminant $\D<0$ implying that 
both the saddle points are complex, which for stable Hartle-Hawking 
no-boundary Universe is seen to be both imaginary. 
These are given by
\beq
\label{eq:pt_sads_12_HH}
\tilde{p}_1 = - 3 i \sqrt{k} \, ,
\hspace{5mm}
\tilde{p}_2 = \frac{3i \sqrt{k} (x-2)}{x} \, .
\eeq
This shows that while saddle point $\tilde{p}_1$ remains fixed at 
the same position for all $x$, the saddle point $\tilde{p}_2$ moves 
from negative imaginary axis to positive imaginary axis 
as $x$ increases. It becomes zero for $x=2$. 
At the saddle point $\tilde{p}_1$, we have a vanishing initial on-shell 
geometry, i.e., $\bar{q}_i=0$, while at the other saddle point $\tilde{p}_2$, 
we have a non-vanishing initial geometry $\bar{q}_i = 12 k (x-1)/\Lam x^2$. The `on-shell'
value of $B(\tilde{p})$ ($B(\tilde{p})$ computed at the 
saddle points) is given by
\beq
\label{eq:Bsad_onshell_HH}
B(\tilde{p}_1) = -\frac{6i k^{3/2}}{\Lam} \, ,
\hspace{5mm}
B(\tilde{p}_2) = -\frac{6i k^{3/2}}{\Lam} \left(
\frac{2}{x^3} - \frac{6}{x^2} + \frac{6}{x} -1
\right) \, .
\eeq
It should be highlighted that for stable Hartle-Hawking no-boundary
Universe referring to $P_i = -3 i \sqrt{k}$ the `action' 
$B(\tilde{p})$ given in eq. (\ref{eq:argExp_Pt_phi}) is complex. 
The morse function $h(\tilde{p})$ at the saddle points for all values of $x$ is given by
\beq
\label{eq:Morse_sad_Bact_xL1}
h(\tilde{p}_1) = \frac{6k^{3/2}}{\Lam} \, ,
\hspace{5mm}
h(\tilde{p}_2) = \frac{6k^{3/2}}{\Lam} 
\left(
\frac{2}{x^3} - \frac{6}{x^2} + \frac{6}{x} -1
\right) \, .
\eeq
Higher derivatives of $B(\tilde{p})$ with respect to $\tilde{p}$
at the saddle point is given by,
\beq
\label{eq:dubDerB_pt_sad}
\biggl. \frac{{\rm d}^2 B}{{\rm d} \tilde{p}^2} \biggr|_{\tilde{p} = \tilde{p}_1}
= - \frac{2i (x-1)\sqrt{k}}{x\Lam} \, ,
\hspace{5mm}
\biggl. \frac{{\rm d}^2 B}{{\rm d} \tilde{p}^2} \biggr|_{\tilde{p} = \tilde{p}_2}
= \frac{2i (x-1)\sqrt{k}}{x\Lam} 
\, ,
\hspace{5mm}
\biggl. \frac{{\rm d}^3 B}{{\rm d} \tilde{p}^3} \biggr|_{\tilde{p} = \tilde{p}_{1,2}}
= \frac{2}{3\Lam} \, .
\eeq
One can expand the function $B(\tilde{p})$ around the saddle point $\tilde{p}_\sg$
where $\sg = \{1,2\}$. This gives
\beq
\label{eq:Bexp_ptsig}
B(\tilde{p}) = B(\tilde{p}_\sg)
+ \biggl. \frac{{\rm d} B(\tilde{p})}{{\rm d} \tilde{p}} \biggr|_{\tilde{p}_\sg} \de \tilde{p}
+ \frac{1}{2} \biggl. \frac{{\rm d}^2 B(\tilde{p})}{{\rm d} \tilde{p}^2} \biggr|_{\tilde{p}_\sg} (\de \tilde{p})^2
+  \frac{1}{6} \biggl. \frac{{\rm d}^3 B(\tilde{p})}{{\rm d} \tilde{p}^3} \biggr|_{\tilde{p}_\sg} (\de \tilde{p})^3 \, ,
\eeq
where $\de \tilde{p} = \tilde{p} - \tilde{p}_\sg$. The series stops at cubic order as the
highest power of $\tilde{p}$ in $B(\tilde{p})$ is three. 
The second variation at the saddle-point can be written 
as $B''(\tilde{p}_\sg)= r_\sg e^{i \rho_\sg}$,
where $r_\sg$ and $\rho_\sg$ depends on boundary conditions. 
Near the saddle point the change in $i H$ will go like 
\beq
\label{eq:changeH}
\de (iH) \propto i 
\left(B''(\tilde{p}_\sg) \right) \left(\de \tilde{p} \right)^2
\sim v_\sg^2 e^{i\left(\pi/2 + 2\ta_\sg + \rho_\sg \right)} \, ,
\eeq
where we write $\de \tilde{p} = v_\sg e^{i\ta_\sg}$ and $\ta_\sg$ is the direction of flow lines
at the corresponding saddle point. 
Given that the imaginary part $H$ remains constant along the 
flow lines, so this means 
\beq
\label{eq:flowang}
\ta_\sg = \frac{(2k-1)\pi}{4} - \frac{\rho_\sg}{2} \, ,
\eeq
where $k \in \mathbb{Z}$. 

For the steepest descent and ascent flow lines, their 
corresponding $\ta_\sg^{\rm des/aes}$ is such that the phase for 
$\de H$ correspond to $e^{i\left(\pi/2 + 2\ta_\sg + \rho_\sg \right)} = \mp1$. This implies
\beq
\label{eq:TaDesAes}
\ta_\sg^{\rm des} = k \pi + \frac{\pi}{4} - \frac{\rho_\sg}{2} \, ,
\hspace{5mm}
\ta_\sg^{\rm aes} = k \pi - \frac{\pi}{4} - \frac{\rho_\sg}{2} \, .
\eeq
These angles can be computed numerically in our case.  

Under the saddle point approximation, the contour integral given in 
eq. (\ref{eq:Bint_def_Bact}) can be computed using Picard-Lefschetz methods
(see \cite{Narain:2021bff} for details of PL methodology).
This gives
\bea
\label{eq:Bint_sadPL_apr}
\int_{-\infty}^{\infty} \frac{{\rm d} \tilde{p}}{2\pi\hbar}
e^{i B(\tilde{p})/\hbar}
&&
= \sum_\sg \frac{n_\sg}{2\pi\hbar} e^{i B(\tilde{p}_\sg)/\hbar}
\int_{{\cal J}_\sg} {\rm d} \de\tilde{p}
\exp\left[i \frac{B''(\tilde{p}_\sg)}{2\hbar} (\de \tilde{p})^2 \right] \, ,
\notag \\
&&
= \sum_\sg n_\sg \sqrt{\frac{1}{2\pi\hbar \lvert B''(\tilde{p}_\sg) \rvert}}
e^{i\ta_\sg} e^{i B(\tilde{p}_\sg)/\hbar} \, .
\eea
where ${\cal J}_\sg$ refers to steepest descent line
and $n_\sg$ is the intersection number which will take
values ($0, \pm1$) accounting for the orientation of contour over each thimble.
In the following 
we will be using this expression to compute the contour integral 
for various values of $x$.

\subsection{$0 < x < 1$}
\label{BsadsXless1}

For $0< x<1$, both the saddle points $\tilde{p}_1$ and $\tilde{p}_2$ lie on the negative
imaginary axis with $\lvert \tilde{p}_2 \rvert< \lvert \tilde{p}_1\rvert$, 
thereby implying that $\tilde{p}_2$ lie below $\tilde{p}_1$ on the negative
imaginary axis. 

\begin{figure}[h]
\centerline{
\vspace{0pt}
\centering
\includegraphics[width=4in]{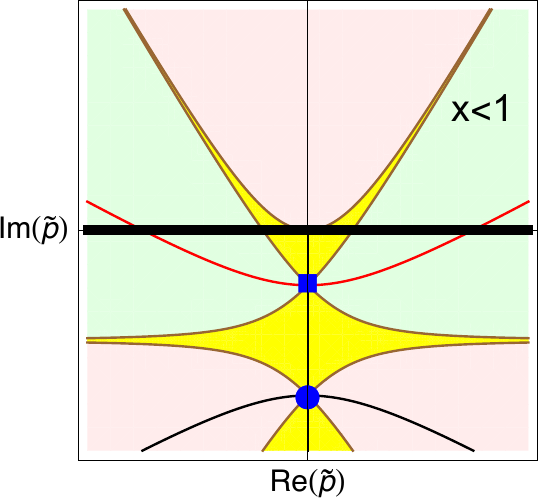}
}
\caption[]{
The gravitational path integral with the Gauss-Bonnet contribution 
favours $P_i=-3i\sqrt{k}$ as the stable configuration 
which also happens to correspond to Hartle-Hawking 
no-boundary Universe. For this value of $P_i$ the action 
for $B$ given in eq. (\ref{eq:argExp_Pt_phi}) is complex.
This figure arises in association with the Picard-Lefschetz analysis of the 
contour integral given in eq. (\ref{eq:Bint_def_Bact}). 
The red lines correspond to the steepest descent lines (thimbles ${\cal J}_\sg$), while 
the thin black lines are the steepest ascent lines and are denoted by ${\cal K}_\sg$. 
Here we choose parameter values: $k=1$, $\Lam=3$, and $x=1/2$. 
For this, the saddle point $\tilde{p}_1$ corresponds to blue-square, while 
saddle point $\tilde{p}_2$ corresponds to blue-circle. Only the 
saddle point $\tilde{p}_1$ (blue square) is relevant. 
The steepest ascent lines emanating from it intersect the 
original integration contour $(-\infty, +\infty)$, which is shown by
thick-black line. The Morse-function $h$ is positive for both saddle points:
$h(\tilde{p}_{1,2})>0$. The light-green region is the allowed region 
with $h<h(\tilde{p}_\sg)$ for all values of $\sg$. The light-pink region 
(forbidden region) has $h>h(\tilde{p}_\sg)$ for all $\sg$. 
The intermediate region is depicted in yellow. 
The boundary of these regions is depicted in brown lines. 
Along these lines, we have $h = h(\tilde{p}_\sg)$.
}
\label{fig:Bactsad_xL1}
\end{figure}

For both the saddle points, the morse function is positive:
$h(\tilde{p}_{1,2})>0$. In figure \ref{fig:Bactsad_xL1}, we plot the various 
flow-line, saddle points, and forbidden/allowed regions. 
From the graph, we notice that only the steepest ascent line from 
$\tilde{p}_1$ intersects the original integration contour, 
thereby making it relevant. At this saddle point, the initial geometry is 
observed to be vanishing, thereby satisfying the ``compactness'' 
and ``regularity" criterion of Hartle-Hawking no-boundary proposal. The thimbles passing through 
this saddle constitute the deformed contour of integration. 
It should also be specified that for $x<1$, the second-derivative 
of $B(\tilde{p})$ computed at saddle point $\tilde{p}_1$ is 
proportional to $+i$, which in saddle point approximation 
gives appropriate Gaussian weight allowing to do Gaussian 
integration. 

The Picard-Lefschetz theory then gives the following  
in the saddle point approximation as
\beq
\label{eq:transamp_PL_xL1}
\int_{-\infty}^{\infty} \frac{{\rm d} \tilde{p}}{2\pi\hbar}
e^{i (P_i - \tilde{p})^2/2 \hbar \bt} \biggl. \Psi_1(\tilde{p}) \biggr|_{\al =0}
\approx \sqrt{\frac{ \Lam x k^{-1/2}}{4\hbar\pi (1-x)} } \,\,
e^{\frac{6 k^{3/2}}{\hbar \Lam}} \, .
\eeq
This regime reproduces the same exponential factor as in Hartle-Hawking
no-boundary Universe. 
This regime leads to stable perturbations 
for the Einstein-Hilbert gravity studied in \cite{DiTucci:2019dji}.
It is expected that Gauss-Bonnet gravitational corrections 
which are topological in nature in 4D won't alter this as long as perturbations are small.

\subsection{$x>1$}
\label{BsadsXmore1}

For $x>1$ both saddles still lie on the imaginary axis. However, only saddle 
point $\tilde{p}_1$ remains fixed on the negative imaginary axis for 
all values of $x$, while the saddle point $\tilde{p}_2$ moves from
negative imaginary axis to positive imaginary axis. It crosses the origin
at $x=2$ in the case of boundary condition corresponding to 
Hartle-Hawking no-boundary Universe. 

\begin{figure}[h]
\centerline{
\vspace{0pt}
\centering
\includegraphics[width=2.7in]{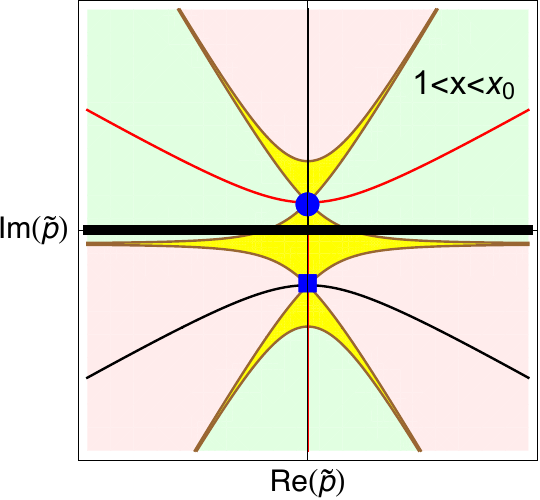}
\hspace{10mm}
\includegraphics[width=2.7in]{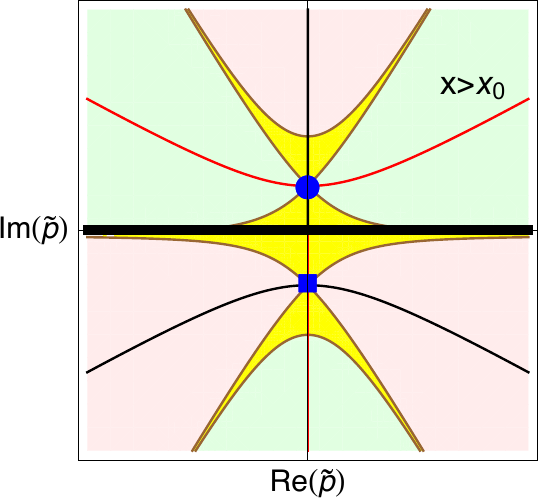}
}
\caption[]{
Here we choose parameter values: $k=1$, $\Lam=3$, and $x=4$ and $x=10$ respectively. 
For this, the saddle point $\tilde{p}_2$ corresponds to blue-circle, while 
saddle point $\tilde{p}_1$ correspond to blue-square. Only the 
saddle point $\tilde{p}_2$ (blue square) is relevant. 
The Morse-function $h$ for $\tilde{p}_1$ is always 
positive, while $h(\tilde{p}_2)$ goes from positive to negative as 
$1<x<\infty$. The crossover happens at $x=x_0$ given by 
eq. (\ref{eq:x_x0_crossover}). 
}
\label{fig:Bactsad_xM1}
\end{figure}

The Morse function corresponding to saddle point $\tilde{p}_1$ is 
always positive, while the Morse function corresponding to 
$\tilde{p}_2$ goes from positive to negative. The crossover 
takes place at $x=x_0$ where $x_0$ is given in 
eq. (\ref{eq:x_x0_crossover}). After the cross-over 
the overall exponential weight becomes negative,
thereby implying an inverse Hartle-Hawking regime.
For $x>1$ it is seen that only the steepest ascent 
curves emanating from $\tilde{p}_2$ intersect the 
original integration contour, implying that it is the 
only relevant saddle point. At this saddle, the initial geometry of the 
universe is non-vanishing, given by $\bar{q}_i= 12 k (x-1)/{\Lam x^2}. $ The thimbles passing 
through this saddle will constitute the deformed contour of integration.
The Picard-Lefschetz analysis then gives the following  
in the saddle point approximation as
\beq
\label{eq:transamp_PL_xM1}
\int_{-\infty}^{\infty} \frac{{\rm d} \tilde{p}}{2\pi\hbar}
e^{i (P_i - \tilde{p})^2/2 \hbar \bt} \biggl. \Psi_1(\tilde{p}) \biggr|_{\al =0}
\approx 
\sqrt{\frac{ \Lam x k^{-1/2}}{4\pi\hbar (x-1)} } \,\,
e^{\frac{6 k^{3/2}}{\hbar \Lam}\left(
\frac{2}{x^3} - \frac{6}{x^2} + \frac{6}{x} -1
\right)} \, .
\eeq
However, the perturbations in this regime are unstable 
for Einstein-Hilbert gravity and the Gauss-Bonnet is expected 
to not change this.

\subsection{$x=1$}
\label{BsadsX=1}

\begin{figure}[h]
\centerline{
\vspace{0pt}
\centering
\includegraphics[width=4in]{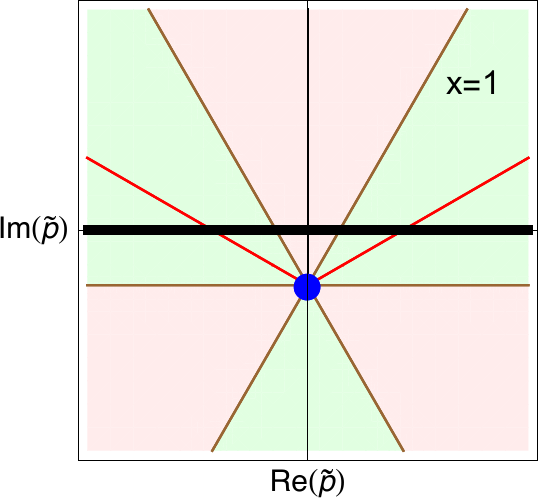}
}
\caption[]{
Here we choose parameter values: $k=1$, $\Lam=3$, and $x=1$. 
This is the degenerate situation for which 
the saddle point $\tilde{p}_1$ and $\tilde{p}_2$ coincide. 
The Morse-function $h$ is positive for both saddle points:
$h(\tilde{p}_{1,2})= 6k^{3/2}/\Lam = 2$. 
}
\label{fig:Bactsad_xeq1}
\end{figure}

This is the degenerate case. The discriminant $\D=0$ for $x=1$ and both 
saddle points coincide : $\tilde{p}_1 = \tilde{p}_2 = -3 i \sqrt{k}$, 
with vanishing initial geometry of the universe at the relevant saddle. 
In this degenerate case both the on-shell action ($B(\tilde{p}_1) = B(\tilde{p}_2)$) 
and the Morse-function become equal ($h(\tilde{p}_1) = h(\tilde{p}_2)
= 6k^{3/2}/\Lam)$. 
In this situation, the saddle-point approximation breaks down 
as the second derivative $B''(\tilde{p}_\sg)=0 $. One needs to go beyond
the second order in the series expansion in 
eq. (\ref{eq:Bexp_ptsig}). In this series, the third order term is non-zero.
This situation is depicted in figure \ref{fig:Bactsad_xeq1}.
In this case, the contour integration gives the following 
\beq
\label{eq:transamp_PL_xDE1}
\int_{-\infty}^{\infty} \frac{{\rm d} \tilde{p}}{2\pi\hbar}
e^{i (P_i - \tilde{p})^2/2 \hbar \bt} \biggl. \Psi_1(\tilde{p}) \biggr|_{\al =0}
\approx 
\frac{\sqrt{3}}{2\pi\hbar} e^{iB(\tilde{p}_\sg)/\hbar} \int_0^\infty {\rm d} v_\sg
e^{-v_\sg^3/9\Lam \hbar}
= \frac{(\Lam /3\hbar^2)^{1/3}}{\Gamma(2/3)}
e^{\frac{6 k^{3/2}}{\hbar \Lam}} \, .
\eeq
%

\section{Conclusions and Outlook}
\label{conc}

In this paper we consider the gravitational path integral of 
Gauss-Bonnet gravity and study it directly in the Lorentzian signature
in four spacetime dimensions. 
Gauss-Bonnet sector of gravity being topological in nature in 
4D doesn't contribute to the bulk dynamics of the field
but has an active role to play at the boundaries or in situations 
where boundaries play an important role.
One such situation is the path integral which is sensitive to 
boundary conditions. Past studies have investigated the effects of 
Gauss-Bonnet sector of gravity on the transition amplitude which is 
given by the gravitational path integral \cite{Narain:2021bff, Narain:2022msz}.
These studies focussed on exploring the consequence of imposing 
Neumann boundary condition at the initial time. 
In this paper we study extensively the effects of imposing Robin boundary 
condition at the initial time and investigate the role played 
by the Gauss-Bonnet sector of gravity systematically in the
Lorentzian gravitational path integral. We setup the platform for raising 
and addressing these issues in the reduced setup of the mini-superspace approximation.

We start by first considering path integrals of a particle in one-dimensional 
potential for various boundary conditions. The path integral 
is evaluated for three different choices of boundary condition: 
Dirichlet boundary condition (DBC) at both initial and final time
(fixing position of particle at two end points), 
Neumann BC (fixing conjugate momenta denoted in paper 
by $p_i$) and Robin BC (fixing linear combination of conjugate 
momenta and position at the initial time: $p_i + \bt q_i = P_i$, where 
$\bt$ is some parameter) with Dirichlet at final time. 
The transition amplitudes are shown to be inter-related with each other
by integral transforms. By exploiting these inter-relations via 
integral transforms one can compute the path integral with NBC or RBC at 
initial time (and DBC at final time) from the path integral of DBC at 
both end points.
These inter-relations can be further manipulated to express the 
RBC path integral as an integral transform of the NBC path integral. 
These results get later utilized in the paper for the computation of 
gravitational path integral in the mini-superspace. 

We take a fresh look at the gravitational path integral in the mini-superspace 
approximation in the Lorentzian signature with Robin boundary 
condition at the initial boundary (that is fixing linear 
combination of conjugate momenta and field
at the initial boundary). Past works have shown that 
NBC and RBC at the initial time leads to stable Universe
\cite{DiTucci:2019bui,Narain:2021bff, Lehners:2021jmv, 
DiTucci:2020weq, DiTucci:2019dji, Narain:2022msz},
while DBC at initial time leads to unsuppressed perturbations
\cite{Feldbrugge:2017fcc, Feldbrugge:2017mbc, Feldbrugge:2017kzv}.
These motivate us to study RBC gravitational path integral 
more carefully for the Gauss-Bonnet gravity 
(the case for Neumann BC was investigated in \cite{Narain:2022msz}). 
The transition amplitude from one 
$3$-geometry to another is given by a path integral 
over $q(t)$ and a contour integration over lapse $N_c$. 

The paper systematically studies the path integral 
in the mini-superspace approximation for the Gauss-Bonnet gravity. 
The gravitational action is varied and carefully analysed to setup a 
consistent variational problem with Robin BC at the initial boundary. 
This process leads to dynamical equation of motion and a 
set of surface terms that need to be supplemented to the action 
to make the gravitational system a consistent variational problem. 
In this process we construct the surface term needed for the 
Gauss-bonnet gravity with the Robin boundary condition.
This surface action is proportional to $\bt$ and smoothly 
reduces to the surface term for the case of 
Neumann BC problem \cite{Narain:2022msz} in the limit of $\bt\to0$. 
To our knowledge this wasn't known in literature earlier.
The path integral is then studied for the total action which has been 
supplemented by these surface terms. 

The path integral to be computed is given in eq. \ref{eq:Gamp},
where $S_{\rm tot}$ for the NBC and RBC case are given in 
eq. (\ref{eq:Sact_frw_simp_nbc}) and (\ref{eq:Sact_frw_simp_RBC})
respectively. To compute the path integral over $q(t)$ 
we make use of the results derived in one-dimensional 
quantum mechanical problem discussed in section \ref{one-part}. 
In the NBC case: the lapse $N_c$ integration and path integral 
over $q(t)$ can be performed exactly as it was also shown in 
\cite{Narain:2022msz}. The numerical prefactor has been correctly 
computed as we do the computation of path integral via first principles 
(method of `time-slicing'). In the RBC case: results of sub-section \ref{rbc_free}
and algebraic manipulations allow us to reduce the problem 
into the product of two Airy-integrals (modulo exponential prefactors). 
One is just the lapse $N_c$ integration and gives rise to 
$\Psi_2(q_f)$ (Airy-integral dependent only on the final 
boundary). The other is an integration over $\tilde{p}$: 
doing a Gaussian integral transform of $\Psi_1(\tilde{p})$ at $\al=0$. 
With a change of integration variable, it is easy to 
convert this into an Airy integral defined as $\Phi(P_i, \bt)$ 
given in eq. (\ref{eq:Phi_Pi_beta}), and an exponential prefactor
dependent on $P_i$ and $\bt$. 
All this when put together gives the exact transition amplitude 
in the RBC case which is given in eq. (\ref{eq:Gbd0bd1_full_rbc}). 
In the limit $\bt\to0$ this correctly reproduces the exact
NBC transition amplitude given in eq. (\ref{eq:Gbd0bd1_full_nbc}). 
This exact transition amplitude for the RBC 
including the Gauss-Bonnet effects is new and hasn't been 
known in literature earlier. 

We then consider the $\hbar\to0$ limit of the exact transition amplitude.
This allows us to single out and focus on configurations giving 
dominant contribution to the path integral. The Gauss-Bonnet 
contribution which appears as an exponential prefactor only,
shows that in the $\hbar\to0$ limit two configurations 
will give dominant contribution: $P_i = -3i\sqrt{k}$ 
and $P_i = 3i\sqrt{k}$. The first one corresponds to 
stable configuration while the later is unstable. 
Coincidentally, $P_i = -3i\sqrt{k}$ also is the configuration 
for the Hartle-Hawking no-boundary Universe where the 
perturbation are well-behaved. In a sense the Gauss-Bonnet
contribution naturally picks and favors the Hartle-Hawking no-boundary 
condition while the other boundary condition is disfavoured. 
Same thing happens for large Gauss-Bonnet coupling 
with $\hbar$ not small. In this case, we are still in deep 
quantum regime however the large Gauss-Bonnet coupling 
favors the Hartle-Hawking no-boundary Universe. This 
is truly a non-perturbative feature. 

We next consider the $\hbar\to0$ limit of the other terms 
in the RBC transition amplitude for the 
case of $P_i = - 3i \sqrt{k}$ (Hartle-Hawking no-boundary Universe). 
This allows us to find the boundary configuration characterised by 
$\bt$ which will give dominant contribution to the RBC transition 
amplitude. It is seen that that $\bt=\bt_{\rm dom} = -i \Lam /2\sqrt{k}$
gives the dominant contribution. On scaling our $\bt = \bt_{\rm dom} x$
we see that the domain of $x$ gets separated into various 
regimes. For $x\leq 1$ we get the same Hartle-Hawking exponential 
prefactor $e^{6 k^{3/2}/\hbar \bar{\Lam}}$ where $\bar{\Lam}$ is 
given in eq(\ref{eq:HHrbcFactor}). For $1<x<x_0$ (where 
$x_0$ is given in eq. (\ref{eq:x_x0_crossover})), the argument
of the exponential prefactor remains positive but monotonically decreases. 
At $x=x_0$ crossover happens and the argument becomes 
negative entering in inverse Hartle-Hawking regime. 
For $x_0<x<\infty$, the argument of exponential decreases 
but asymptotically approaches to $e^{\frac{-6k^{3/2}}{\hbar\Lam}(1-\frac{4\alpha\Lam}{3})}$
which is the exponential prefactor obtained with 
Dirichlet boundary condition at $t=0$ and is known 
to be unstable as perturbations are unsuppressed
\cite{Feldbrugge:2017fcc, Feldbrugge:2017mbc, Feldbrugge:2017kzv}.
This analysis shows that the allowed region where we 
correctly reproduce the Hartle-Hawking exponential 
prefactor is $0\leq x \leq 1$. This was noticed in
\cite{DiTucci:2019dji, DiTucci:2019bui}
via a different route. 

We then study the transition amplitude in the saddle-point 
approximation and make use of Picard-Lefschetz methods 
to compute the contour integrals. However, we address the problem 
in a different manner. Past studies in this direction applied 
Picard-Lefschetz methodology to the lapse $N_c$-integration 
after the $q(t)$ path integral has been worked out \cite{DiTucci:2019dji, DiTucci:2019bui}. 
Here as we managed 
to express RBC transition amplitude as NBC transition amplitude 
via eq. (\ref{eq:Gamp_rbc_ms_exp}), so it gave us the flexibility 
of approaching the problem in another way. Instead of direct lapse $N_c$
integration on the full on-shell action, we do saddle point analysis of only 
the momentum integration given in eq. (\ref{eq:Bint_def_Bact}). 
Saddle analysis of the 
$N_c$-integration acting only on the NBC part was done in 
\cite{Narain:2022msz}. It showed that there are two saddle points for 
all values of $q_f$. For $q_f<3k/\Lam$ the two saddle points lie on 
the imaginary axis in the complex $N_c$ plane (only one is relevant), 
while for $q_f>3k/\Lam$ the two saddles lie on real axis in the 
complex $N_c$ plane (both are relevant). As $q_f$ increases 
Universe undergoes a transition from Euclidean to Lorentzian 
phase \cite{Narain:2022msz}. In this paper, we just study the 
$\tilde{p}$-integral via Picard-Lefschetz methods. 

The saddle analysis of the $\tilde{p}$-integration shows that 
in the complex $\tilde{p}$-plane there are always two 
saddle points, both lying on the imaginary $\tilde{p}$ 
axis. The saddle point $\tilde{p}_1$ is independent of 
$x$ (remains fixed), with vanishing initial geometry, while the 
saddle point $\tilde{p}_2$ varies with respect to $x$, with non-vanishing initial geometry,
and gets pushed to infinity (on the negative imaginary axis) as $x\to0$ (Neumann limit).
For all values of $x$, only one of the two saddle point
is relevant. For $0 < x <1$ the saddle point $\tilde{p}_1$ 
relevant, while the irrelevant saddle point $\tilde{p}_2$
lies below $\tilde{p}_1$ on the negative imaginary axis 
in the complex $\tilde{p}$ plane (in the limit $x\to0$ this saddle point
is pushed to infinity and remains irrelevant). 
In this regime, we get the 
correct Hartle-Hawking exponential prefactor
$e^{6 k^{3/2}/\hbar \bar{\Lam}}$. For $x>1$, the saddle 
$\tilde{p}_2$ becomes relevant. 
$x=1$ is the degenerate situation when $\tilde{p}_1=\tilde{p}_2$ (relevant)
giving $e^{6 k^{3/2}/\hbar \bar{\Lam}}$ (Hartle-Hawking) beside the numerical
pre-factors.
For $1<x<2$ the saddle 
point $\tilde{p}_2$ still lies on the negative imaginary axis.
However, for $x>2$ it crosses over and lies on the positive 
imaginary axis. For $1<x<x_0$, the argument of the 
exponential prefactor decreases in magnitude and becomes 
negative for $x>x_0$. The range of $x$ for which only the 
saddle point $\tilde{p}_1$ is relevant and produces the 
exact exponential prefactor $e^{6 k^{3/2}/\hbar \bar{\Lam}}$ of Hartle-Hawking 
no-boundary Universe is $0 < x<1$. 

Our investigations show the important non-trivial role played by the 
Gauss-Bonnet sector of the gravitational action in favoring 
the initial configurations which lead to 
the Hartle-Hawking no-boundary Universe. 
Gauss-Bonnet term arises in the low 
Energy-effective action of the heterotic string theory 
\cite{Zwiebach:1985uq,Gross:1986mw,Metsaev:1987zx}
with $\al>0$. Although it is topological in nature in four spacetime
dimensions and is expected to not play any role in the 
dynamical evolution, but our analysis clearly shows
its contribution in the path integral. When Gauss-Bonnet term
is not ignored in path-integral studies, then for $\al>0$
(which is also the same sign appearing in low
energy effective action of the heterotic string theory)
it naturally picks and favors initial configurations which 
correspond to Hartle-Hawking no-boundary Universe. 
Moreover, it is expected that Gauss-Bonnet modifications 
won't alter the stability analysis done for Einstein-Hilbert 
gravity \cite{DiTucci:2019dji}, 
where the perturbative stable regime was found to be
$0 \leq x \leq 1$. This is because Gauss-Bonnet in 
four spacetime dimensions is topological. Perturbative 
analysis like the one done in \cite{DiTucci:2019dji}
is for small perturbations which are expected to not 
change the background topology,
and hence will be unaffected by the Gauss-Bonnet gravity.

\bigskip
\centerline{\bf Acknowledgements} 

We are thankful to Romesh Kaul for various illuminating discussions 
during the course of this work. We would also 
like to thank Chethan Krishnan 
for useful discussions at various stages of the work. 
We also thank the anonymous referee for the helpful suggestions. 

%


\end{document}